\documentclass{article}
\usepackage[margin=1in]{geometry}

\usepackage{amsmath}
\usepackage{amssymb}
\usepackage{graphicx}
\usepackage{enumitem}
\usepackage{booktabs}
\usepackage{array}
\usepackage{tabularx}
\usepackage{xcolor}
\usepackage{subcaption}
\usepackage{authblk}
\usepackage{float}
\usepackage{placeins}

\newcolumntype{Y}{>{\raggedright\arraybackslash}X}

\newtheorem{definition}{Definition}[section]

\newtheorem{observation}[definition]{Observation}

\newcommand{\sig}{\sigma}

\title{Domain Specific Post Quantum Signatures for Blockchains}
\author[1]{Maja Lie}
\author[1,2]{Benjamin Marsh}
\affil[1]{Sei Labs}
\affil[2]{University of Portsmouth}
\date{July 2026}

\begin{document}
\maketitle

\begin{abstract}
Blockchains need more than post quantum single signer signatures. They need
consensus profiled authentication objects with canonical bytes, priced invalid input
rejection, stable transaction identifiers, hybrid downgrade resistance, public
aggregation, merge semantics, accountable signer evidence, forward secure committee
rotation, and light client consequences. We argue for domain specific
post quantum signatures for blockchain roles, analogously to how hash function
engineering produced domain specific primitives for hash table DoS and
arithmetized proof systems. We formalize transaction authorization and
quorum certificate requirements, instantiate them on Bitcoin, Ethereum, and a
Sei Giga style high throughput BFT stress profile, and evaluate ML-DSA,
SLH-DSA, Falcon/FN-DSA, HAWK, MAYO, SNOVA, UOV/QR-UOV, FAEST, SQIsign,
LaBRADOR Falcon, Squirrel, Chipmunk, and DKKW/LeanSig. We find that NIST single signer signatures are necessary components, yet none of the current
schemes is a sufficient drop in replacement for the signature layer of modern public
blockchains. The missing object is a consensus ready post quantum signature
profile, not another generic size table.
\end{abstract}

\section{Introduction}
\label{sec_intro}

A blockchain is a shared, decentralized digital ledger that records transactions. Digital signatures have two main functions within a blockchain: the first, to prove that a transaction was authorized by a particular user (transaction authorization), and the second, to attest that participating validators have verified and approved a proposed block (quorum certificate). In addition to these functionalities, the size and verification time of signatures is important to the scalability and performance of a blockchain. This poses a challenge for blockchain migration to post quantum signatures which are significantly larger and slower to verify than classical signatures such as ECDSA. 

The usual comparison table for post quantum signatures has three columns, namely signature size,
verification time, and nominal NIST security category. Although these columns are relevant, they
do not form a blockchain functionality specification. For example, a blockchain verifier is a public service
that must process adversarially chosen invalid inputs, a transaction signature is part of a
consensus byte string, a validator signature is often consumed as part of a
quorum certificate rather than individually, and slashing requires public evidence about which validators signed which
conflicting messages. In this setting, BLS~\cite{cfrg-bls-signature-07} serves as more than a signature by providing an operational
interface with public aggregation, compact certificates, cheap verification, signer descriptors,
and mergeable vote gossip. Replacing BLS therefore means replacing that interface, not merely
choosing a post quantum single signer primitive.

None of the proposed post quantum signature schemes currently support such an interface. NIST finalized ML-DSA and SLH-DSA as FIPS 204 and FIPS 205 in August 2024~\cite{nist2024standards,fips204,fips205}. Falcon was selected for standardization as FN-DSA, but as of May 2026 public IETF material still describes FIPS 206 as expected in late 2026 or early 2027 rather than as a finalized FIPS~\cite{ietf-cose-falcon}. Blockchain deployments have nevertheless begun experimenting with Falcon style verification. For example, Algorand exposes \texttt{falcon\_verify} as an AVM v12 opcode using the compressed Falcon signature format, making this an on chain verification primitive rather than, by itself, a statement that Falcon is the native account signature scheme \cite{algorand-falcon-brief,algorand-falcon-verify}.

The gap identified in this paper goes beyond the size and speed of post quantum signatures. The deeper issue is that most academic post quantum signature work
targets the wrong abstraction for public blockchains. The standard abstraction is a
single signer primitive whose verifier returns accept or reject. A blockchain signature
is instead a consensus object with adversarial byte parsing, transaction identifier
semantics, fee metering, fork binding and role binding, hybrid downgrade resistance, gossip merge
behaviour, accountable evidence, validator rotation, and light client consequences. A
construction that proves EUF-CMA security but leaves these properties to the application layer
has not solved the blockchain problem. It has only supplied one component that a consensus
profile may or may not be able to use.

Cryptographic standards of practice already support this kind of demand. The
community already accepts domain specific primitives when the execution environment changes.
SipHash is a short input keyed hash/PRF designed for hash table and network facing denial of
service settings rather than as a universal replacement for SHA-2 or SHA-3. MiMC,
Rescue Prime, Poseidon, and related arithmetization oriented hash functions exist because
ordinary bit oriented hashes are badly shaped for SNARK, STARK, MPC, and FHE circuits
\cite{siphash,mimc,poseidon,rescue-prime}. Blockchains create an analogous design pressure for
signatures. Consensus does not merely ask for ``a secure signature.'' It asks for an
authentication object with canonical bytes, priced rejection, stable transaction names,
publicly mergeable vote aggregation, slashable signer evidence, light client verification, and
migration semantics. It would therefore be surprising if the right long term answer were limited
to importing TLS oriented single signer post quantum signatures unchanged.

The research agenda should be stated plainly. Blockchains need domain specific post quantum
signature schemes and profiles. Some will be strict profiles around standardized primitives,
such as ML-DSA, SLH-DSA, or FN-DSA. Others should be new constructions whose native interface
is a transaction, a validator vote, a quorum certificate, a slashing proof, or a light client
update. Treating NIST signatures as leaves inside a larger consensus signature design is the
signature analogue of using SHA-2 as a component while designing Poseidon like hashes for
arithmetized proof systems.

This paper makes four claims.

\begin{enumerate}[leftmargin=*]
\item Blockchain post quantum signature migration should be specified as a pair of
 functionalities, not as one generic signature API. Transaction authorization and validator
 quorum certificates have different adversaries, cost models, and evidence requirements.

\item The requirements should be byte level, cost level, and deployment budgeted. It is not
 enough to say that a scheme is EUF-CMA secure and fast on valid inputs. The consensus
 profile must specify exactly which public keys and signatures are accepted, how they are
 domain separated, how transaction identifiers are computed, which hybrid policy is required,
 and what worst case cost an adversary can impose before rejection.

\item Chain architecture changes which post quantum gap is dominant. Bitcoin is mostly a
 transaction authorization migration problem, Ethereum is both a transaction and BLS style
 aggregation problem, Tendermint style BFT chains can tolerate linear certificates only at
 modest validator counts and block rates, Algorand shows that on chain Falcon verification
 and smart signature overlays are useful but do not remove the need for native account and
 consensus signing profiles, Sui exposes both user signature agility and BLS authority
 certificates, Solana today is an MTU limited transaction signature problem, and
 next generation high throughput EVM/BFT chains make transaction signature bytes and fast QC
 mergeability first order bottlenecks.

\item Current post quantum proposals occupy different
 parts of the design space as components rather than drop in replacements. Among NIST finalized schemes,
 ML-DSA is a plausible transaction primitive if size is acceptable, SLH-DSA is conservative
 but too large for high rate transaction lanes, and Falcon/FN-DSA is compact but demands a
 strict byte and implementation profile. Among onramp candidates, HAWK relaxes Falcon's
 floating point signing burden but rests on younger lattice assumptions, MAYO offers a
 compact multivariate alternative with current security analysis still evolving, SNOVA's
 original size profile was attractive but the submitted Round 2 parameter sets are now in
 flux after recent attacks, and SQIsign offers very compact keys and signatures at signing speeds
 incompatible with high throughput use. Among quorum certificate proposals,
 LaBRADOR style Falcon aggregation gives succinct aggregate verification but is not
 algebraically incremental, synchronized lattice multi signatures (Squirrel, Chipmunk) are
 incremental within a bounded time horizon and require synchronized state, and the recent
 hash based multi signature line of Drake, Khovratovich, Kudinov and
 Wagner~\cite{drake-hashbased-mulsig,leansig} reproduces more of the BLS interface than any
 lattice or SNARK proposal known to us, at a different size and verification cost tradeoff.
\end{enumerate}

\paragraph{Scope.}
The focus of this paper is signature functionality. We do not attempt a complete migration plan covering
state commitments, light client proofs, peer to peer authentication, encrypted mempool
protocols, or post quantum VRFs, although we identify VRFs as an adjacent open problem for
sortition based systems. We also separate functionality from implementation hardening.
Side channel and fault resistance are mentioned where they change the suitability of a
primitive for signing roles, but a complete implementation security treatment is out of scope.
Our evaluation focuses on schemes that are NIST finalized, NIST onramp Round 2 candidates, or
have peer reviewed post quantum aggregation literature. We emphasize the candidates whose
public key, signature size, verification, or aggregation profile is most
directly relevant to blockchain transaction or quorum certificate use. Eliminated Round 2 candidates are discussed only when they remain useful as historical examples of deployment geometry or cryptanalytic risk. We do not survey every public key
signature proposal and we deliberately do not give a new signature construction.

\paragraph{Related work.}
Several existing surveys cover adjacent ground. The exotic signatures SoK~\cite{exotic-sigs-sok}
taxonomizes post quantum threshold, ring, blind, and aggregate signatures with attention to
consensus applications, but does not formalize functional requirements at the byte or cost
level. Performance oriented studies measure ML-DSA, SLH-DSA, and Falcon/FN-DSA verification
on the EVM and other constrained verifiers~\cite{poqeth,bench-blockchain-pqc}. NIST's IR
8528~\cite{nist-ir-8528} reports on the first round of the additional digital signatures
process and is the authoritative current source on many onramp candidates. Cloudflare
provides a useful TLS anchored comparison of the lattice and onramp candidates~\cite{cloudflare-pq-sigs}. Anchor work on blockchain specific aggregate signatures includes the lattice
multi signature line of Squirrel~\cite{squirrel} and Chipmunk~\cite{chipmunk}, the LaBRADOR
aggregation of Falcon by Aardal, Aranha, Boudgoust, Kolby and Takahashi~\cite{aardal-labrador},
and the hash based multi signature framework of Drake, Khovratovich, Kudinov and
Wagner~\cite{drake-hashbased-mulsig}. Recent IETF work on hybrid signature spectra and
strong unforgeable hybrid combiners is directly relevant to blockchain migration profiles
because hybrid authorization is normally the first deployment phase~\cite{ietf-hybrid-sig-spectrum,cfrg-suf-hybrid-sigs}. None of these works provide the requirements/critique structure we develop here.

\paragraph{Organization of the paper.} In Section \ref{sec_core_requirements}, we give a
blockchain facing functionality specification, including the requirements for transaction authorization signatures and signatures for consensus. In Section \ref{sec_req_compiler_compact}, we provide a compiler which, given the functional requirements of a given blockchain, outputs which technical requirements the chain should ask of a signature scheme. Section \ref{sec_sizes} reviews the current design space for post-quantum signatures and gives further details byte/time budget classes for transaction signatures in blockchains. Section \ref{sec_evaluation} evaluates a number of post-quantum signatures against these requirements and Section \ref{sec_case_studies} contextualizes this in real examples of blockchains, showing where single signer standardization is insufficient.

\section{Core transaction and quorum certificate requirements}
\label{sec_core_requirements}

The purpose of this section is to outline the qualities signatures must satisfy to prevent them from being reused in a blockchain in a different context than intended. We start with the first functionality, transaction authorization, then give the requirements for the second functionality, consensus. A blockchain profile fixes an algorithm identifier, public key and signature codecs, length caps, transcript binding, gas/precharge rules, and negative test vectors. 

\subsection{Transaction authorization requirements}\label{sec:tx_sig}

Messages within a blockchain carry specific information like gas fees, which are part of a transaction's signed payload and determine the cost and execution constraints of the transaction. We will refer to requirements for transaction authorization signatures as T class requirements. 

For a transaction, the signed transcript should at least bind chain ID, fork version, role, algorithm suite, public key hash, account or credential mode, nonce, fee fields, and the transaction intent. Binding these fields ensures that a signature authorizes a specific transaction in a specific protocol context and cannot be validly reused or reinterpreted elsewhere. For example, the chain ID and fork version prevent replay across different networks or protocol versions, the role and algorithm suite provide domain separation so signatures cannot be repurposed for other functions, and the nonce prevents replay of the same transaction on the same chain and ensures each transaction is processed at most once. Likewise, binding the fee fields, public key hash, account or credential mode, and transaction intent ensures that the signer explicitly authorizes the execution parameters, authentication context, and intended action, preventing post-signing modification. 

These requirements are covered by T2 in Table \ref{tab_t_reqs_compact}. In addition to this, Table \ref{tab_t_reqs_compact} gives other basic minimum requirements for transaction signatures. Beyond unforgeability, we require canonical encodings and deterministic verification to guarantee consensus consistency across implementations, bounded verification cost and economic precharging to mitigate denial-of-service attacks, reproducible signing for secure wallet and custody deployments, resistance to transaction malleability, and support for the throughput, latency, and resource constraints of modern blockchain networks. T12 through T16 address deployment considerations such as credential placement, which determines where signatures and public keys are stored or referenced within transactions; account abstraction, which enables flexible authentication policies beyond simple externally owned accounts; encrypted mempools, which hide transaction contents before they are processed to improve transaction privacy and reduce opportunities for transaction manipulation; batch verification, which improves throughput by verifying multiple signatures simultaneously; and memory and state budget rules, which bound the storage and computational resources consumed by transaction authorization to ensure predictable performance and scalability.

\begin{table}[!htbp]
\centering
\scriptsize
\setlength{\tabcolsep}{2.2pt}
\renewcommand{\arraystretch}{1.06}
\begin{tabularx}{\textwidth}{lY}
\toprule
\textbf{Req.} & \textbf{Transaction authorization requirement} \\
\midrule
T1/T1a & Multi user post quantum unforgeability with a concrete report for number of active keys, lifetime signing queries, proof loss, hash model, and chain lifetime. \\
T2 & Context and role binding. No transaction signature can replay as a vote, bridge attestation, governance message, or signature on another fork. \\
T3 to T6 & Canonical public key/signature bytes, deterministic verification, total public key validation, and consensus identical rejection of malformed encodings. \\
T4/T11/T17 & Adversarial worst case verification cost and economic precharge. Malformed keys and signatures must be priced before the attacker can externalize CPU or memory cost. \\
T7 & Seeded signing reproducibility for wallets, HSMs, MPC custody, and deterministic audit. This is where Falcon/FN-DSA signing profiles and HAWK implementation flags matter. \\
T8 & Transaction identifier malleability resistance. Either the stable transaction name excludes randomized authorization bytes, or the authorization profile proves strong malleability resistance. \\
T9 to T10 & Byte and honest verification budgets at the chain's target throughput, since a scheme can pass T1 and still fail T9/T10. \\
T12 to T16 & Credential placement, account abstraction, encrypted mempool, batch verification, and memory/state budget rules. \\
\bottomrule
\end{tabularx}
\caption{T class requirements. These are the requirements Bitcoin like, Solana like, EVM account, and high throughput transaction lanes actually need.}
\label{tab_t_reqs_compact}
\end{table}

\subsection{Quorum certificate requirements}

We next consider signatures which get used in consensus to authenticate votes that collectively determine the next block to be committed to the blockchain. Committee signatures are normally consumed as a quorum certificate rather than as $n$ independent authorizations.  A BLS aggregate signature is popular in this role not merely because it is short, but because it gives a particular interface: public aggregation, incremental updates as votes arrive, a signer descriptor, and cheap verification of the final certificate. 

We call requirements relevant to this type of signature as Q class requirements. For a vote profile in consensus, the signed transcript must also bind epoch, height, round, step, block identifier, committee root, and signer weight policy. These fields collectively define the context in which a consensus vote is valid. Including them in the signed transcript ensures that a vote is uniquely bound to a particular epoch, block proposal, and consensus round, preventing replay, and equivocation across protocol contexts. Including the signer weight policy and committee root also ensures that votes are verified against the correct validator set and voting weights for that consensus instance.

\begin{table}[!htbp]
\centering
\scriptsize
\setlength{\tabcolsep}{2.2pt}
\renewcommand{\arraystretch}{1.06}
\begin{tabularx}{\textwidth}{lY}
\toprule
\textbf{Req.} & \textbf{Quorum certificate requirement} \\
\midrule
Q1/Q1$'$ & Quorum soundness, and forward secure quorum soundness when validators use key evolution or stateful leaves. \\
Q2 & Public, non interactive aggregation of independently broadcast votes. Threshold signing alone does not satisfy this. \\
Q3/Q8/Q9 & Unique signer set, weight, committee root, epoch, and transcript binding, rogue key resilience, deterministic verification. \\
Q4/Q11/Q12 & Concrete sublinear cryptographic verification, absolute certificate size budget, and deployed aggregate verification time budget. \\
Q5 & Fresh add, disjoint merge, and overlap safe merge semantics, or an explicit vote cache protocol that supplies gossip idempotence. \\
Q6 & Accountable slashing evidence, full certificates, retained component signatures, or local openings must identify which validator signed conflicting transcripts. \\
Q7/Q10 & Committee rotation and forward security without trusted setup or parameter retuning at every epoch. \\
Q13 to Q18 & Negative aggregate vectors, prover/aggregator cost, aggregator DoS, certificate batching, memory/evidence retention, and PQ soundness of any SNARK/folding/zkVM compression layer. \\
\bottomrule
\end{tabularx}
\caption{Q class requirements. These are the requirements hidden by the phrase ``replace BLS''.}
\label{tab_q_reqs_compact}
\end{table}

The Q class requirements above ensure that aggregated consensus signatures remain secure, efficient, and accountable under realistic blockchain operating conditions. In particular, quorum soundness ensures a quorum certificate can only be produced if the required set of validators genuinely signed the same consensus transcript, while forward-secure soundness preserves this guarantee even if long-term signing keys are later compromised. We also ask for public, non-interactive aggregation, deterministic verification, resistance to rogue-key attacks, and efficient verification and certificate sizes to support high-throughput consensus. 

In addition to being non interactive, aggregation should satisfy safe aggregation semantics, which define how partial quorum certificates can be incrementally built, merged, and deduplicated as votes are exchanged across the network. They also include requirements for accountability, committee evolution, and practical implementation, such as accountable slashing evidence, which enables validators that sign conflicting consensus messages to be identified and penalized; committee rotation and forward security, which allow validator membership and signing keys to evolve securely over time; and negative aggregate test vectors, resilience to denial-of-service attacks on aggregators, certificate batching, memory and evidence retention policies, and post-quantum secure compression techniques for scalable certificate verification.

\section{A compact compiler}
\label{sec_req_compiler_compact}

The previous sections introduced a complete set of blockchain signature requirements. These requirements should not be interpreted as a flat checklist for every signature primitive; instead, the relevant requirements depend on the signature-facing functionality exposed by the blockchain. To make this mapping explicit, we provide a compact requirement compiler which, upon inputting a desired blockchain feature, outputs the corresponding requirements to be considered. One can use this compiler to identify the requirements associated with a desired feature and, in turn, evaluate or select a signature scheme that satisfies those requirements.

The T axis captures native transaction authorization, the Q axis captures validator quorum certificates, and the remaining axes—S, L, V, X, and H—
represent inherited subsystem requirements for stateful signing, light clients, verifiable random functions (VRFs) for cryptographic sortition (e.g., selecting validators or block proposers), cross-chain messaging (bridges), and hybrid migration, respectively. We next summarize the requirements for the additional features associated to the S, L, V, X and H axes.

\paragraph{Stateful signing.} Stateful signatures are attractive in blockchain consensus because hash-based
and key-evolving constructions can buy conservative assumptions and forward
security. However, it is important to be rigorous in the management of signing state or risk destroying the security property the construction was chosen to provide. We give seven S class requirements. S1 requires each one-time or few-time signing state to be uniquely identified and consumed at most once, preventing state reuse that could compromise security. S2 requires mechanisms to prevent or detect state rollback following crashes, backups, or system recovery, ensuring previously consumed signing states cannot be reused. S3 defines safe crash recovery procedures, favouring the loss of uncertain signing capacity over the risk of reusing signing state. S4 specifies the evidence that must be retained to support accountability and slashing throughout the required retention period. S5 requires distributed and threshold signing protocols to coordinate state consumption so that concurrent signers cannot inadvertently use the same signing state. S6 requires explicit reporting of the storage requirements for live state, backups, audit logs, and retained evidence to facilitate practical deployment. S7 requires comprehensive test vectors that validate both correct state transitions and failure scenarios, ensuring implementations correctly handle crashes, duplicate-use detection, and other state-related faults.

\paragraph{Light clients.} A light client is a blockchain client that verifies the blockchain without downloading or storing the entire blockchain state. Instead, it relies on compact cryptographic proofs provided by full nodes. The light client requirements ensure that resource-constrained clients can securely verify blockchain state using compact cryptographic proofs while remaining consistent with full nodes. L1 requires light clients to accept and reject exactly the same certificates, public key encodings, and other consensus-critical data as full nodes, preventing inconsistencies that could compromise consensus safety. L2 requires signed headers to bind both the application state root and the validator committee root, ensuring that state proofs are authenticated by the consensus that produced them. L3 requires reporting the computational and storage costs of verifying light client updates, particularly when verification occurs within another blockchain or virtual machine. L4 specifies update frequency, data retention, and pruning policies so that light clients can continue to verify the blockchain correctly despite changing validator sets or key material. L5 requires secure handling of signature scheme migration, ensuring that historical headers, state proofs, and validator keys remain verifiable without ambiguity during transitions between cryptographic schemes.

\paragraph{Sortition and VRFs.} Sortition-based chains use publicly verifiable randomness to decide who may propose, vote, or form a committee. The VRF requirements ensure that verifiable random functions provide secure, unbiased, and efficient randomness for blockchain sortition. V1 requires VRF outputs to be unique, publicly verifiable, and deterministically validated, preventing ambiguity or manipulation during eligibility checks. V2 requires the output to remain pseudorandom until revealed and binds the proof to the relevant blockchain context, preventing validators from increasing their probability of selection by grinding over keys, randomness, or protocol parameters. V3 requires eligibility proofs to be bound to the correct stake snapshot, delegation state, and committee selection rules so that validator eligibility is determined consistently. V4 requires stateful VRF constructions to satisfy the stateful signing requirements, preventing rollback or reuse of signing state that could enable equivocation about eligibility. V5 requires proof sizes and verification costs to remain within practical resource budgets, ensuring that sortition credentials can be efficiently disseminated and verified within the blockchain protocol.

\paragraph{Cross-chain messaging.} The cross-chain messaging requirements ensure that messages exchanged between independent blockchains remain authentic, unambiguous, and secure under different trust models. X1 requires cross-chain authorizations to be bound to the source and destination chains, communication route, message contents, and finality policy, preventing messages from being replayed or redirected to unintended destinations. X2 requires the verification model used by the destination chain to be explicitly specified, ensuring that the appropriate security requirements are inherited for light clients, committee attestations, threshold custody, or optimistic protocols. X3 requires messages to be bound to the source chain's finality mechanism so that only finalized state can be relied upon by the destination chain. X4 requires destination chains to account for the computational cost of verifying both valid and malformed cross-chain messages, protecting bridge infrastructure from denial-of-service attacks. X5 requires all operational bridge keys to be identified together with their cryptographic protection mechanisms and migration strategy, ensuring that critical bridge operations remain secure throughout the transition to post-quantum cryptography.

\paragraph{Hybrid PQ migration.} The hybrid migration requirements ensure that classical and post-quantum signature schemes can coexist securely during the transition to post-quantum cryptography. H1 requires hybrid authorizations to verify both the classical and post-quantum signatures, ensuring that security depends on the successful verification of both components rather than either individually. H2 requires both signatures to be bound to the same blockchain context and declared suite policy, preventing downgrade attacks or the reuse of hybrid signatures as classical-only or post-quantum-only authorizations. H3 specifies how the component signatures interact, either by explicitly preventing their independent reuse or by defining a policy that ensures they cannot be misused outside the intended hybrid authorization. H4 requires deterministic verification procedures and consensus-defined resource accounting so that all nodes process hybrid signatures consistently and remain resilient to malformed inputs. H5 requires transaction identifiers to be derived from the signed transaction intent rather than randomized signature bytes, preventing transaction malleability and ensuring stable transaction identifiers. H6 requires explicit migration policies specifying when classical-only signatures are no longer accepted, ensuring that hybrid signatures serve as a temporary transition mechanism rather than a permanent compatibility layer.

\begin{table}[!htbp]
\centering
\scriptsize
\setlength{\tabcolsep}{2.5pt}
\renewcommand{\arraystretch}{1.06}
\begin{tabularx}{\textwidth}{lYY}
\toprule
\textbf{Blockchain feature} & \textbf{Requirements emitted} & \textbf{Reason} \\
\midrule
User transaction authorization & T1 to T12, T16 to T17, add T13 for account abstraction, T14 for encrypted mempools, T15 for batch verification, H1 to H6 during migration & Public adversaries choose keys, messages, encodings, and invalid signatures, while fee and mempool rules must price rejection before execution. \\
Validator voting & Q1 to Q3, Q6 to Q10, Q13, Q17, add S1 to S7 if votes are stateful or key evolving & A vote is also a consensus object used in fork choice, finality, and slashing. \\
Quorum certificates & Q1 to Q18 & A BLS like certificate needs public aggregation, signer descriptors, weight binding, merge semantics, accountability, and light client friendly verification. \\
Light clients and bridges & L1 to L5, X1 to X5, and whichever T/Q/H requirements protect the source object & A bridge or light client verifies another chain's signature object under a stricter gas and calldata budget. \\
Sortition / VRF systems & V1 to V5 plus S1 to S7 if credentials are stateful & PQ signatures do not automatically supply PQ uniqueness, anti grinding, and stake bound randomness. \\
Hybrid migration & H1 to H6 attached to every migrated role & AND semantics, suite binding, gas charging, txid policy, and sunset rules are consensus rules, not wallet preferences. \\
\bottomrule
\end{tabularx}
\caption{Compact requirement compiler from chain functionality to signature requirements.}
\label{tab_req_compiler_compact}
\end{table}

\paragraph{Compiled examples.} In the rest of this paper, we consider three flavors of blockchains based on real chains such as: Bitcoin, Ethereum, and Sei Giga. Bitcoin and Ethereum represent the two dominant blockchain architectures, while Sei Giga serves as a representative of emerging multiple concurrent proposer (MCP) blockchain designs, which allows multiple validators to propose blocks simultaneously, increasing parallelism and improving throughput. Migrating a Bitcoin-style blockchain, where transactions spend previously created outputs rather than update account balances, primarily introduces T and H requirements. It also requires consistent parsing of transaction signatures, clear policies for transaction identification and size accounting, secure handling of existing outputs protected by legacy signatures, and explicit rules for phasing out classical cryptography. Ethereum-like proof of stake chains emit both T and Q class requirements, account authorization, account abstraction, BLS style
attestation replacement, sync committee/light client verification, and
forward secure validator migration. A Sei Giga-style high throughput BFT
profile has the harshest requirements, including below one kilobyte transaction signatures and low latency accountable quorum certificates with fresh add and merge semantics, allowing independently collected validator votes to be incrementally combined into the same quorum certificate as they propagate through the network.

\section{Sizes, budget classes, and the design space}
\label{sec_sizes}

Before evaluating individual schemes against the requirements, it is worth seeing the design
space at a glance. Table~\ref{tab_sizes} collects public key, signature, and approximate
verification cost figures for the schemes we consider, taken from the relevant
specifications and round 2 submission packages where available. We include pre quantum ECDSA
and BLS rows for calibration. All sizes are byte level profile sizes, not merely mathematical
object sizes, where a standard or submission admits multiple encodings, the row assumes the
encoding named in the notes column. Approximate rows are meant to show geometry, not to serve
as a benchmark.

\begin{table}[H]
\centering
\small
\setlength{\tabcolsep}{3.2pt}
\renewcommand{\arraystretch}{1.15}
\resizebox{\textwidth}{!}{%
\begin{tabular}{llrrrl}
\toprule
\textbf{Scheme} & \textbf{status/cat.} & \textbf{pk (B)} & \textbf{sig/cert (B)} & \textbf{verify} & \textbf{notes} \\
\midrule
ECDSA secp256k1 & classical & 33 & 71 & fast & pre quantum reference \\
BLS12-381, Ethereum/Sui style profiles & classical & 48/96 & 96/48 & fast & pre quantum reference, group choice profile dependent \\
\midrule
ML-DSA-44 & FIPS 204 / I & 1\,312 & 2\,420 & fast & finalized \cite{fips204} \\
ML-DSA-65 & FIPS 204 / III & 1\,952 & 3\,309 & fast & finalized \cite{fips204} \\
SLH-DSA-128f & FIPS 205 / I & 32 & 17\,088 & slow & finalized \cite{fips205} \\
SLH-DSA-128s & FIPS 205 / I & 32 & 7\,856 & slow & finalized \cite{fips205} \\
Falcon-512 / FN-DSA & selected / I & 897 & 666 & fast & padded, FIPS 206 pending \cite{falcon-spec,ietf-cose-falcon} \\
Falcon-1024 / FN-DSA & selected / V & 1\,793 & 1\,280 & fast & padded, FIPS 206 pending \cite{falcon-spec,ietf-cose-falcon} \\
\midrule
HAWK-512 & onramp R2 / I & 1\,024 & 555 & fast & \cite{hawk-spec} \\
HAWK-1024 & onramp R2 / V & 2\,440 & 1\,221 & fast & \cite{hawk-spec} \\
MAYO-1 / MAYO-2 & onramp R2 / I & 1\,420 / 4\,912 & 454 / 186 & fast & compact signature tradeoff \cite{mayo-spec} \\
UOV classic, level I & onramp R2 / I & $\sim$272\,000 & 96 & fast & very large pk, compressed variants smaller \cite{uov-spec} \\
SNOVA, submitted R2 & onramp R2 / in flux & 1\,016 & 248 & fast & attractive sizes, R2 params attacked \cite{snova-spec,snova-wedge-attack} \\
QR-UOV & onramp R2 / I & $\sim$12\,000 to 14\,000 & $\sim$200 & moderate & compressed UOV variant \\
CROSS & onramp R2 / I & $<121$ & $\sim$9\,000 to 13\,000 & moderate & code based \cite{cross-spec} \\
SQIsign-I & onramp R2 / I & 65 & 148 & moderate & compact, slow signing \cite{sqisign-spec} \\
FAEST (level I, EM) & onramp R2 / I & 32 & $\sim$5\,000 & slow & VOLE in the head \cite{faest-spec} \\
\midrule
\multicolumn{6}{l}{\emph{Aggregate / multi signature constructions.}} \\
BLS aggregate ($|S|$=128) & classical & n/a & 96 & fast & aggregate certificate, pre quantum \\
LaBRADOR Falcon ($|S|$=128) & PQ / research & n/a & $\sim$60\,000 & sublinear & SNARK style certificate \cite{aardal-labrador} \\
Squirrel ($|S|$=1024) & PQ / research & n/a & $\sim$36\,000 & moderate & synchronized, bounded \cite{squirrel} \\
Chipmunk ($|S|$=1024) & PQ / research & n/a & $\sim$20\,000 & moderate & synchronized, bounded \cite{chipmunk} \\
DKKW/LeanSig individual vote & PQ / research & n/a & $\sim$2\,000 to 5\,000 & n/a & per validator signature, not aggregate \cite{drake-hashbased-mulsig,leansig} \\
DKKW/LeanSig aggregate QC & PQ / research & n/a & tens of kB & sublinear & aggregate certificate, parameters vary \cite{drake-hashbased-mulsig,leansig} \\
\bottomrule
\end{tabular}%
}
\caption{Public key, signature, and indicative verification cost figures for the schemes discussed. Sizes are approximate where parameter sets vary, and readers should consult the specifications for exact values. ``fast'' means sub millisecond verification on commodity x86, ``moderate'' means 1 to 10 ms, and ``slow'' means $\geq 10$ ms. Aggregate construction rows distinguish per validator vote signatures from aggregate quorum certificate sizes. The table is intended to show design space geometry rather than serve as a rigorous benchmark.}
\label{tab_sizes}
\end{table}

The geometry is plain. At NIST level 1, all the post quantum lattice schemes for
single shot signing land in the hundreds of bytes to 2.5 KB range with sub millisecond
verification. In particular, SLH-DSA is much larger, but has conservative security guarantees. Younger schemes such as multivariate ones (MAYO, SNOVA like designs) continue to have fast moving security analyses. The multivariate family makes the
opposite tradeoff to the lattice family, that is, signatures are very small, but public keys are either
huge (UOV) or land in the lattice public key range. Isogeny based SQIsign has the smallest combined
public key plus signature size of any candidate but pays in signing and verification time. In the aggregate
setting, BLS sets a target that no post quantum proposal currently matches across the full
requirement set, although the recent hash based multi signature line approaches the operational
interface most closely.

\subsection{Named blockchain budget classes}
\label{sec_named_budgets}

The previous requirements are profile parametric, but a purely parametric statement hides the
engineering reality. We therefore use the following named budget classes as reference points.
They are falsifiable deployment targets rather than standards.

\begin{table}[H]
\centering
\caption{Reference transaction signature budget classes. TMTU and THT are deliberately severe because they are the classes in which ML-DSA and SLH-DSA cease to be plausible default transaction signatures.}
\scriptsize
\setlength{\tabcolsep}{3pt}
\renewcommand{\arraystretch}{1.08}
\begin{tabularx}{\textwidth}{lrrrrY}
\toprule
\textbf{Class} & \textbf{sig budget} & \textbf{pk in tx budget} & \textbf{honest verify} & \textbf{invalid reject} & \textbf{intended use} \\
\midrule
TMTU & $\le 512$ B & $\le 96$ B & $\le 100\,\mu$s & $\le 2\times$ valid or prepaid worst case & Solana style packet limited transactions \\
THT & $\le 1024$ B & usually 0 & $\le 50$ to $100\,\mu$s & $\le 2\times$ valid or prepaid worst case & high throughput EVM/BFT transaction lane \\
TGEN & $\le 3$ KB & $\le 2$ KB & $\le 1$ ms & explicit gas bound & ordinary account based chains \\
TROOT & $\le 20$ KB & flexible & $\le 10$ ms & explicit gas bound & governance, roots, emergency keys \\
\bottomrule
\end{tabularx}
\subcaption*{TMTU = packet-constrained transaction profile;\\
THT = high-throughput transaction profile;\\
TGEN = general-purpose transaction profile;\\
TROOT = low-frequency root, governance, or emergency authorization profile.}
\label{tab_tx_budget_classes}
\end{table}

These budget classes provide concrete deployment targets for signatures in blockchain applications. While security requirements determine whether a scheme is cryptographically suitable, the budget classes capture the operational constraints imposed by blockchain architectures and the services they are designed to provide, including limits on signature size, public key size, verification latency, and the cost of rejecting malformed inputs. These constraints vary considerably across use cases, from latency-sensitive, high-throughput transaction processing to infrequent governance operations where larger signatures are acceptable.

\begin{table}[H]
\centering
\caption{Reference quorum certificate budget classes. QBLS like is currently not met by any post quantum construction known to us. QLC is the class in which DKKW/LeanSig style constructions become plausible. QBFT100 is the class in which linear certificates may still be tolerable.}
\scriptsize
\setlength{\tabcolsep}{3pt}
\renewcommand{\arraystretch}{1.08}
\begin{tabularx}{\textwidth}{lrrrrY}
\toprule
\textbf{Class} & \textbf{committee} & \textbf{QC budget} & \textbf{AggVerify} & \textbf{Add/Merge} & \textbf{intended use} \\
\midrule
QBLS like & $128$ to $2048$ & $\le 1$ KB & $\le 2$ ms & $\le 100\,\mu$s & direct BLS replacement for light clients and hot consensus \\
QLC & $128$ to $2048$ & $\le 32$ KB & $\le 10$ ms & explicit, gossip safe & light client friendly PQ certificates \\
QBFT100 & $\le 100$ & $\le 256$ KB & $\le 20$ ms & cache assisted acceptable & modest validator BFT chains \\
QHT & $100$ to $512$ & $\le 16$ KB & $\le 2$ ms & $\le 100\,\mu$s & next generation high throughput BFT \\
\bottomrule
\end{tabularx}
\subcaption*{
QBLS-like = BLS-like quorum-certificate profile;\\
QLC = light-client-oriented quorum-certificate profile;\\
QBFT100 = quorum-certificate profile for BFT committees of at most 100 validators;\\
QHT = high-throughput quorum-certificate profile.
}
\label{tab_qc_budget_classes}
\end{table}

The reference budgets illustrate the challenge of deploying current post-quantum signature schemes in blockchain systems. Signature budgets ranging from a few hundred bytes to a few kilobytes align poorly with many post-quantum constructions, whose signatures often range from around one kilobyte to tens of kilobytes depending on the underlying cryptographic family and security level. 

\subsection{Adversarial verification measurements}
\label{sec_awcv_measurements}

A blockchain verifier cannot be evaluated solely by its ability to accept valid signatures; it must also reject malformed and adversarial inputs efficiently and consistently. This is essential because blockchain nodes verify data from untrusted users, and inconsistent or excessively expensive rejection of malformed inputs can threaten both consensus safety and network availability. A deployment profile must report adversarial verification cost on invalid inputs. The adversarial worst case verification (AWCV) and negative test vector requirements turn this into a concrete artifact. The lack of such data is itself a deployment blocker for variable length or compressed encodings. 

\begin{table}[H]
\centering
\scriptsize
\setlength{\tabcolsep}{3pt}
\renewcommand{\arraystretch}{1.08}
\begin{tabularx}{\textwidth}{lYYY}
\toprule
\textbf{Profile} & \textbf{Encoding} & \textbf{Structural AWCV character} &
\textbf{Measurement gap and deployment consequence} \\
\midrule
ML-DSA-44 strict
 & fixed length, bit packed
 & parsing constant in byte length, verification dominated by fixed degree
 NTT and norm checks, AWCV structurally tight against honest path cost.
 & honest path numbers exist in the NIST submission and PQClean, a consensus
 profile must publish negative vector AWCV measurements to confirm the
 structural expectation. Tightly meterable in principle. \\

Falcon-512 padded only
 & padded fixed length, salt plus compressed Gaussian polynomial
 & decompression is the worst case parsing path, length maximal malformed
 signatures can exercise the full Huffman style unary decoding before the
 verifier rejects.
 & maximum decompression cost on length maximal random and structured invalid
 inputs must be measured. Acceptable in a high throughput profile only if
 the chain precharges decompression cost at the padded length before
 decoding. \\

Falcon-512 permissive
 & padded, unpadded, and uncompressed variants all accepted
 & each accepted encoding has its own parsing path, one gas calibration
 cannot cover all three without precharging the maximum across encodings.
 & per encoding AWCV plus encoding switch overhead. Should fail any native
 consensus profile unless every accepted encoding is independently priced,
 the safer fix is to restrict acceptance to a single encoding. \\

HAWK-512 strict
 & padded compressed signature, integer only verifier
 & structurally similar to Falcon, compressed decoding dominates the worst
 case, verification itself is fixed cost integer arithmetic once decoded.
 & adversarial parsing measurements on compressed inputs and
 outside distribution coefficients. No consensus profiled AWCV measurements
 are known to the authors at the time of writing. \\

MAYO-1
 & fixed length signature, multivariate map carried in the public key
 & signature side parsing bounded by finite field operations on fixed
 dimensions, adversarial worst case concentrates in public key validation
 and structured invalid map evaluation.
 & malformed public key parsing cost and worst case map evaluation under
 structured invalid signatures. The public key validation rule must be
 stated as part of the profile, not deferred to a verifier library. \\
\bottomrule
\end{tabularx}
\caption{Structural AWCV characterization for candidate transaction signature
profiles.}
\label{tab_awcv_structural}
\end{table}

\paragraph{Artifact boundary.}
This paper does not claim to supply empirical AWCV measurements. The AWCV
table above is a structural classification of where worst case rejection cost
is expected to arise, fixed length parsing, compressed decoding, public key
validation, wrong context rejection, or aggregate descriptor rejection. Each entry states what the parsing and verification structure of the profile implies about adversarial worst case cost, distinguished from the empirical measurement obligations of the AWCV measurement suite. ``Tightly meterable'' and ``structurally similar to Falcon'' are statements about parsing structure, not substitutes for measured upper bounds. A proper deployment profile still needs a public measurement artifact with the following
shape,
\begin{align*}
    \text{profile} \quad &\text{valid median} \quad \text{random invalid median} \quad \text{malformed max} \\&\qquad \text{near valid max} \qquad \text{max/valid}
\end{align*}
reported for the input families of the AWCV measurement suite. The matrices in this paper should therefore be read as classifying \emph{meterability} and missing measurement obligations, not as reporting completed gas calibration
benchmarks. This distinction is important: a scheme can be structurally
meterable and still not be consensus ready until the negative vector corpus and
worst case rejection measurements exist. Moreover, concrete median, 95th, 99th, and tail rejection costs over the input families of the AWCV measurement suite still need to be published before a profile is treated as consensus ready. 

\section{Evaluation of current proposals}
\label{sec_evaluation}

In this section, we look at a number of current proposals, including those recommended for standardization by NIST and those still undergoing evaluation. Proposals which have not been selected to progress to round 3 of NIST's call for additional signatures are still included where they are relevant as case studies of deployment constraints or necessary to highlight cryptanalytic vulnerabilities. For each construction or family of schemes, we give the relevant size and cost figures from Tables~\ref{tab_sizes}, \ref{tab_tx_budget_classes}, and \ref{tab_qc_budget_classes}, and assess against the requirements of the transaction profile (T1 to T17) or the QC profile (Q1 to Q18). Two general remarks apply throughout.
First, all entries are profile statements rather than statements about a raw mathematical scheme, a scheme that fails a requirement under a permissive byte profile may pass under a strict one, and we say so. Second, where the open literature does not yet support a definitive judgement, we mark the entry as partial and explain. We are deliberately harsher here than a standards track primitive survey would be. A primitive can be cryptographically valuable and still fail the blockchain interface.

\subsection{ML-DSA (FIPS 204)}
\label{sec_mldsa}

ML-DSA is the cleanest transaction signature candidate among the finalized NIST lattice schemes. Signatures and public keys have fixed size byte encodings, signing and verification use only integer arithmetic, and the verifier rejects malformed inputs deterministically with
modest variance. Its primary drawback is size, signatures are 2\,420, 3\,309, and 4\,627 bytes for ML-DSA-44/65/87 respectively \cite{fips204}.

Against the transaction profile, ML-DSA satisfies requirements T1--T7, with its principal limitation being the communication overhead imposed by its signature size. T1 (multi user post quantum unforgeability) rests on ML-DSA's standardized security claim and supporting analysis, lifted to the multi user setting only after a concrete multi target budget is stated. T3 (canonical byte acceptance) is straightforward because ML-DSA signatures are encoded with fixed length and are bit packed with explicit padding that the verifier rejects on
mismatch. T4 (adversarially metered) is plausible because parsing is constant time in the byte length and verification is constant time in the number of NTT operations once parsing succeeds. T5 (consensus deterministic) holds under a fixed FIPS compatible profile. T7 (seeded signing
reproducibility) is achievable because ML-DSA signing is integer only and the FIPS 204 specification fixes the rejection sampling and randomness derivation, deterministic and randomized signing modes are both specified. T2 (context and role binding) is achieved at the protocol level rather than by ML-DSA itself: the blockchain profile binds the chain identifier, fork version, signer role, and other contextual information into the signed transcript, and ML-DSA authenticates that transcript. Similarly, T8 (transaction-identifier non-malleability) is a chain profile decision, the safe default is to compute the transaction identifier from the signed intent and exclude signature bytes. If
a chain insists on including ML-DSA signature bytes in the transaction identifier, it should
state the strong unforgeability or malleability resistance argument it relies on. FIPS 204 also makes length checks part of the security sensitive profile, a blockchain verifier should not treat truncated or extended encodings as harmless library inputs.

The remaining work is profile work rather than algorithm selection. That work covers concrete multi user bounds, exact transcript binding, invalid input metering, negative test vectors, and canonical byte tests. The hard failure is with the budget classes (TMTU/THT) for transaction signatures. ML-DSA-44's 2.4 KB signature is acceptable for many conventional blockchain platforms, but it is too large for systems with strict transaction size limits and becomes costly in high throughput networks designed to process hundreds of thousands of transactions per second. A high throughput chain can use ML-DSA for root keys, account upgrade certificates, governance, bridge administration, or fallback authorization, but it is not a drop in replacement for compact ECDSA/Schnorr transaction authorization at line rate.

As for fitting the second functionality, quorum certificates, ML-DSA does not provide a native public aggregate signature interface comparable to BLS\@.
The committee setting therefore requires either a separate Q class construction
(Sections~\ref{sec_lattice_multisig}, \ref{sec_snark_agg}, and \ref{sec_hashbased_multisig}), a SNARK or folding layer
over per signer ML-DSA signatures, or willingness to verify $|S|$ individual signatures. The first option inherits whichever Q class limitations the chosen aggregate has, the second option pays the SNARK style limitations we discuss in Section~\ref{sec_snark_agg}, the third option scales linearly with committee size and breaks Q4 above some committee size.

\paragraph{Verdict.} ML-DSA is the safest finalized transaction signature primitive in the post quantum portfolio, subject to a chain's tolerance for multi kilobyte signatures. It fits the transaction budget class for TGEN better than TMTU or THT. It does not, by itself, solve the quorum certificate problem.

\subsection{SLH-DSA (FIPS 205)}
\label{sec_slhdsa}

SLH-DSA is hash based, stateless, and the most conservative of the finalized NIST signatures,
its security rests on the second preimage and pseudorandomness assumptions of the underlying
hash function alone. It passes T1 to T7 naturally under a strict profile, fixed encoding,
hash only verification, no floating point, deterministic mode available, at a substantial
size cost. T8 is again a transaction identifier policy question, not a property to be left
implicit. SLH-DSA-128f signatures are 17\,088 bytes, the ``s'' parameter sets bring this down
to roughly 7\,856 bytes at the cost of significantly slower signing \cite{fips205}.

For high throughput transaction lanes, SLH-DSA is impractical, at 17 KB per signature, even a
1\,000 transaction block carries 17 MB of signature payload, before transactions themselves are
counted. The natural deployment is in low frequency, high value roles, root keys, governance,
emergency rotation, and roots of trust for stateful hash based signatures further down a
hierarchy. In these roles, its conservatism is an asset.

\paragraph{Verdict.}
SLH-DSA is the right primitive for low frequency, high assurance roles. It is the wrong
primitive for transaction rate signing on any high throughput chain and should be marked as
failing budget classes TMTU, THT, and most ordinary hot path transaction profiles.

\subsection{Falcon / FN-DSA}
\label{sec_falcon}

Falcon's attraction is obvious, at NIST level 1, padded Falcon-512 signatures are 666 bytes
and Falcon-1024 signatures are 1\,280 bytes, much smaller than ML-DSA at comparable security
categories \cite{falcon-spec,ietf-cose-falcon}. Verification is integer only in the Falcon
design, and NIST's FN-DSA material likewise treats verification as not requiring
floating point arithmetic \cite{nist-fips206-status}. This makes Falcon appealing as a
\emph{verification} primitive for blockchain settings.

Four issues require care in the transaction profile.

\paragraph{T3 (canonical byte acceptance) is profile dependent.}
Falcon's signature contains a salt and a Huffman style compressed Gaussian polynomial. The
specification defines a padded encoding of fixed length, and Falcon's decompression algorithm
rejects negative zero coefficients and nonzero trailing bits. However, the specification also
describes an unpadded variable length encoding and an alternate uncompressed encoding, and
verifiers \emph{may} support these alternative formats \cite{falcon-spec}. A consensus profile
must therefore choose exactly one accepted format and reject all others. A natural choice is
to accept only the fixed length padded form for the selected parameter set, with exact
public key encoding and fixed transcript binding. Under this strict profile, T3 holds. Under
a permissive profile, T3 is at best partial and may fail native consensus readiness.

\paragraph{T4 and T11 (adversarially metered) are measurement dependent.}
Falcon's variable length compressed representation uses a unary like component in
decompression, so malformed inputs can exercise parsing paths whose costs are not captured by
average case valid signature benchmarks. With a fixed maximum length and a verifier that
charges for worst case parsing before decoding, this is meterable in the sense of
the metered verification definition. Under permissive support for unpadded or multiple encodings, the cost
model becomes substantially messier and the simple ``charge for average valid verification''
gas schedule undercharges adversarial inputs. A blockchain profile that wants Falcon in a
transaction lane should publish the T11 negative vector table before deployment rather than after an
invalid input denial-of-service incident.

\paragraph{T7 (seeded signing reproducibility) is structurally hard.}
According to NIST's public FN-DSA status material, signing relies on native or emulated IEEE-754 binary64 floating-point arithmetic, implementations are required to exactly reproduce the published known-answer tests, and the draft specification constrains the order of operations while prohibiting fused multiply-add instructions. The status update also indicates randomized signing
only and warns that deterministic signing modes can be dangerous under floating point mistakes.
The practical consequence for a chain is that Falcon/FN-DSA can be profiled for deterministic
verification but seeded signing reproducibility is an implementation obligation rather than a
structural property. A consensus deployment that needs cross platform reproducibility must
mandate a single reference signing implementation, including its specific FFI bindings, and
treat divergence as a chain fork rule. This is a real operational cost worth naming
explicitly as it converts the choice of FN-DSA from a selection of an algorithm into a selection
of a particular implementation profile.

\paragraph{T8 and strong unforgeability are not automatic.}
Falcon is randomized. Randomized signing by itself is acceptable, although a chain must not rely
on unique signature bytes for transaction identity unless it has a strong malleability resistance
argument. The clean profile is to identify transactions by signed intent and treat signature
bytes as authorization evidence rather than as part of the stable transaction name. The
recent Falcon proof discussion around salt, public key binding, and tight multi user security
underscores that a blockchain profile should not silently assume the exact proof properties it
wants unless the deployed profile matches the proved variant.

\paragraph{Implementation attack surface.}
Independently of the consensus profile, the BEARZ work reports a fault attack analysis of
Falcon's trapdoor sampler and simulated private key recovery from the faulted
computation \cite{bearz}. Single trace power analysis on Falcon-512 signing on commodity
microcontrollers has also been demonstrated. Neither result makes Falcon unusable, but they
argue against using Falcon signing casually on commodity hardware without a hardened
implementation. This is more relevant for the validator role, where signing happens on
heterogeneous hardware, than for transaction signing in professionally managed custody.

\paragraph{Verdict.}
Falcon/FN-DSA can be profiled for safe blockchain transaction verification under a strict
encoding policy and a defined gas schedule for invalid input rejection. Its signing side
operational requirements are a real cost that should be priced in before committing to it as
the chain default. Falcon is a plausible THT candidate only under strict profiling, a
permissive Falcon profile should be treated as failing native consensus readiness.

\subsection{HAWK}
\label{sec_hawk}

HAWK is a NIST onramp Round 3 lattice signature designed by the same group as
Falcon \cite{hawk-spec}. It uses a hash and sign structure but its trapdoor sampler is
integer only, sidestepping the floating point operational issue that dominates Falcon. Sizes
are competitive with Falcon, HAWK-512 has a 1\,024 byte public key and a 555 byte signature,
HAWK-1024 has 2\,440 byte and 1\,221 byte respectively. The transaction profile evaluation is mostly favorable, with a few caveats.

\paragraph{Security assumption maturity.}
HAWK's security argument relies on two problems, the one more shortest vector problem
(omSVP) and search module lattice isomorphism (smLIP), that are variants of more
conventional lattice problems but have received less cryptanalytic attention than the
M-LWE/M-SIS assumptions underlying ML-DSA and the NTRU based assumptions underlying Falcon.
NIST's IR 8528 explicitly notes this and encourages further study \cite{nist-ir-8528}. This
does not critique the construction. It notes that T1 currently rests on a younger
foundation in HAWK than in ML-DSA or Falcon. In the functional matrix we mark T1 as satisfied
under HAWK's stated assumptions and put the maturity issue in the deployment matrix.

\paragraph{T7 partially recovered.}
HAWK signing is integer only in principle, which removes the cross architecture floating point
hazard. However, the reference implementation's optimization flag \verb|HAWK_FMA| enables use
of FMA compiler intrinsics for performance, and the implementation documentation explicitly
notes that this flag can change generated keys and signatures from a given seed, breaking
test vector reproducibility \cite{hawk-impl}. A consensus profile that mandates HAWK without
FMA recovers T7, one that permits implementation choice loses it. HAWK also uses compressed
encodings, so the canonical encoding and AWCV considerations from Falcon apply analogously.

\paragraph{AWCV is not optional.}
Because HAWK's blockchain attraction is compactness and fast verification, its deployment case
is precisely the TMTU/THT setting where invalid input metering matters most. A HAWK profile
without negative vector parsing data should not be called consensus ready, even if the valid
signature benchmark is excellent.

\paragraph{Side channel exposure.}
Power analysis attacks on the unprotected HAWK reference implementation have been
demonstrated \cite{hawk-side-channel}. As with Falcon, this is an implementation hardening
question rather than a structural disqualification, but it is part of the operational cost of
HAWK signing on physical devices.

\paragraph{Verdict.}
HAWK is the natural ``Falcon but with integer signing'' candidate. Under a strict profile that
fixes the implementation flags and the byte format, HAWK is a strong transaction primitive.
Adopters should treat the assumption maturity question and missing AWCV data as the dominant
risk factors, not the size cost tradeoff.

\subsection{Multivariate schemes including UOV, MAYO, SNOVA, and QR-UOV}
\label{sec_multivariate}

The multivariate family rests on the difficulty of solving systems of multivariate quadratic
equations over a finite field. It is the only family in NIST's onramp that consistently offers
small signatures with non lattice security assumptions. The size profile is inverted relative
to lattice signatures: signatures are small while public keys are often large. The security history
profile is also different, the family has attractive long running ideas, but several prominent
structured multivariate signatures have been badly broken. The right blockchain posture is
therefore to separate functional fit from current cryptanalytic confidence.

\paragraph{UOV.}
Classical UOV gives 96 byte signatures with roughly 272 KB public keys at NIST level
1 \cite{uov-spec}. This is not a T6 failure, public key validation can still be deterministic
and metered. The problem is deployment fit. A 272 KB public key per active account is
prohibitive for on chain account storage, and embedding the public key in every transaction is
also prohibitive. The UOV submission includes smaller key variants, for example around 43 KB
at level 1, at the cost of slower verification \cite{uov-spec}, that still does not naturally
match the usual blockchain account model.

\paragraph{MAYO.}
MAYO is a UOV variant that retains UOV style small signatures while shrinking the public key
into the lattice range \cite{mayo-spec}. At NIST level 1, the Round 2 profile gives several
operating points, including MAYO-1 with a 1\,420 byte public key and a 454 byte signature, and
MAYO-2 with a 4\,912 byte public key and a 186 byte signature. This makes MAYO the most
credible multivariate candidate for a blockchain transaction profile, with sizes broadly
comparable to Falcon or better on signatures. The transaction profile holds up
qualitatively, integer arithmetic, fixed encodings, deterministic verification, subject to
two reservations. First, the multivariate security argument is younger and has a recent track
record of broken proposals, most notably Rainbow. Second, the verification cost profile of
multivariate schemes against adversarially malformed inputs is less studied in the consensus
literature than the lattice family, AWCV bounds for MAYO are not, to our knowledge, published.

\paragraph{SNOVA.}
SNOVA is another UOV variant whose original size profile was excellent, signatures around 248
bytes and public keys around 1\,016 bytes at level 1 \cite{snova-spec}. That profile should no
longer be presented as deployable without qualification. Recent cryptanalysis exploiting
SNOVA's block ring structure reports that the updated Round 2 parameters are broken, bringing
SNOVA-I down to about 94 bits of security under the attack model \cite{snova-wedge-attack}.
This leaves room for future SNOVA like constructions, while making the submitted Round 2 SNOVA
profile a moving target rather than a current blockchain transaction signature candidate.

\paragraph{QR-UOV.}
QR-UOV reduces UOV's public key size to the 12 to 14 KB range at the cost of a more involved
verification procedure. At this public key size it remains structurally awkward for blockchain
accounts, and its verification cost is reported as needing further optimization in NIST's
status reporting \cite{nist-ir-8528}.

\paragraph{Verdict for the family.}
MAYO is the multivariate candidate worth watching for transaction signing because it offers compact
signatures with public keys small enough to fit many account models, gated on the maturity of
multivariate security analysis and AWCV measurements. UOV and QR-UOV have public key profiles
that fit the usual blockchain account model poorly. SNOVA had attractive byte geometry, although
the submitted Round 2 parameters should not be treated as deployable after current attacks.
None of these constructions provide a native aggregate signature interface, so the QC role is
unaddressed. Despite several recent attacks, NIST has moved all multivariate proposals to Round 3 \cite{nist-round-3}. 

\subsection{Code based and MPCitH style schemes including CROSS, FAEST, MQOM, and SDitH}
\label{sec_codebased}

\paragraph{CROSS.}
CROSS is a code based onramp Round 2 candidate. Public keys are very small, but signatures are
large, and verification is moderate speed. This is a SLH DSA shaped tradeoff with a code based
assumption rather than a hash based one. Like SLH-DSA, CROSS is plausible for low frequency
high assurance roles where signature size is not the bottleneck. Both code based proposals, LESS and CROSS, from Round 2 of NIST's call for additional signatures were dropped from Round 3 due to uncertainty in their security following attacks in the second round \cite{nist-round-3}.

\paragraph{FAEST.}
FAEST builds on the VOLE in the head paradigm to obtain a signature whose security is built
around AES like symmetric assumptions \cite{faest-spec}. Signatures are roughly 5 KB at level 1
with roughly 32 byte public keys. FAEST is structurally interesting because it relies on a
long studied symmetric primitive with a different assumption profile from lattices,
multivariate equations, codes, and isogenies. It is inaccurate to say that every blockchain
already trusts AES in its hash chain, since many chains rely on SHA-2, Keccak, BLAKE, Poseidon,
Rescue, or other primitives instead. The right point is assumption diversification rather than
universal AES deployment. Verification is slow relative to the lattice family and signature
sizes are mid sized. Its fit for blockchain transactions is therefore similar to SLH-DSA,
low frequency high assurance, not throughput.

CROSS and FAEST receive shorter treatment because their blockchain fit is structurally clear,
while their cryptographic interest remains real. They diversify assumptions and are
plausible for low frequency authorization, but neither supplies compact transaction rate
signatures nor a QC aggregation interface. From the perspective of this paper, they are
important as conservative fallback components, not as candidates for the missing BLS like
post quantum consensus object.

\paragraph{MQOM.}
MQOM advanced to Round 3 and gives a different point in the size/assumption
space, small public keys, signatures in the low kilobyte range at level I, and
an MQ/MPCitH style security foundation \cite{mqom-spec,nist-round-3}. From a
blockchain perspective this puts MQOM between the compact lattice/multivariate
transaction candidates and the larger assumption diversification candidates.
Its public keys are account model friendly, but the signatures are too large
for TMTU and generally too large for THT. MQOM is therefore worth tracking
for TGEN, TROOT, governance, bridge administration, and low frequency
authorization, but it does not address Q class aggregation.

\paragraph{SDitH.}
SDitH also advanced to Round 3. It uses a syndrome decoding in the head
construction with small public keys and level I signatures in the several kB
range \cite{sdith-spec,nist-round-3}. Its blockchain fit is similar to FAEST
and MQOM rather than Falcon, HAWK, or MAYO. It diversifies assumptions and can
serve low frequency authorization roles, but it is not a compact transaction
primitive for packet limited or very high throughput chains. Like the other
single signer candidates in this subsection, SDitH supplies no native BLS like QC
interface.

\paragraph{Verdict.}
CROSS is now best treated as a historical Round 2 code based candidate rather
than an active NIST onramp candidate. FAEST, MQOM, and SDitH remain active
Round 3 candidates and are valuable assumption diversification primitives, but
their current size and verification profiles make them better fits for
low frequency authorization than for TMTU, THT, or QBLS like profiles. None
of them addresses Q class aggregation.

\subsection{Isogeny based, SQIsign}
\label{sec_sqisign}

SQIsign is the isogeny based onramp Round 2 candidate and has advanced to Round 3 \cite{sqisign-spec}. At NIST level 1 the
Round 2 public key is 65 bytes and the signature is 148 bytes, smaller than any other
post quantum signature by a substantial margin and within a small constant factor of
pre quantum ECDSA. Verification is in the single digit millisecond range on commodity x86,
signing is much slower. The Round 2 SQIsign documentation reports roughly 103 Mcycles for
level I signing and 5.1 Mcycles for verification in the optimized implementation, which is
dramatically improved over earlier SQIsign variants but still leaves signing far outside a
high throughput hot path \cite{sqisign-spec}. The security assumption, the supersingular
endomorphism ring problem, diversifies NIST's portfolio, which is otherwise lattice dominated.

For a blockchain transaction profile, SQIsign's size is attractive but its signing latency is
not. Slow signing is incompatible with high throughput user experience, even for occasional
high value transactions it is a meaningful cost. SQIsign signing also appears to be hard to
implement in a timing side channel secure manner \cite{cloudflare-pq-sigs}, which is an
implementation hardening cost. T1 holds under the isogeny assumptions, T3 to T6 hold
qualitatively, T7 is unaddressed in the current specification, and T8 should be handled by
intent based transaction identifiers unless the chain has a separate malleability resistance proof.

\paragraph{Verdict.}
SQIsign is the right primitive for compactness critical low frequency roles, light client
proofs, archived attestations, settlement signatures over rollup batches, where slow signing
is acceptable. It is not a transaction signature for a high throughput chain.

\subsection{Quorum certificate baseline, BLS}
\label{sec_bls}

For comparison purposes we summarize what BLS aggregate signatures provide, since they are the
de facto interface baseline against which post quantum QC proposals are measured \cite{bgls03}.
BLS gives a 96 byte aggregate signature in the common Ethereum BLS12-381 profile regardless of
the number of signers. Some systems, including Sui authority signatures, choose the dual
minSig profile with 48 byte signatures and 96 byte public keys \cite{sui-auth-overview}. In
both cases, aggregate verification is dominated by a multi pairing, descriptor processing and
aggregate public key computation are still linear or authenticated linear in the signer
descriptor, but the cryptographic work replaces $|S|$ individual signature verifications with
one aggregate verification. Aggregation is public and non interactive. Adding a fresh signer
is one group multiplication, and disjoint aggregates can be merged cheaply.

Two caveats are important. First, bare BLS aggregate bytes do not support arbitrary
overlap idempotent merge, if two partial aggregates share a signer, naive multiplication
duplicates that signer's contribution. Real consensus protocols obtain gossip idempotence from
signer descriptors, per signer vote caches, and deduplication before aggregation. Second,
conflicting BLS aggregate certificates do not generally let one reconstruct individual
signatures from aggregate bytes. Slashing can be handled by accepting the full certificates
plus canonical signer descriptors as computational evidence, or by requiring retention of
individual votes during the slashing window. It should not be described as aggregating again the
complement of the intersection unless the protocol actually retains the decomposable material
required to do so.

Thus BLS is the correct operational baseline, but the baseline is a \emph{protocol profile},
not merely the 96 byte or 48 byte aggregate signature object.

\subsection{Forward secure and key evolving constructions}
\label{sec_fs_qc}

Forward security is a separate axis from the multi signature/SNARK/hash based taxonomy used in
Sections~\ref{sec_bls}, \ref{sec_lattice_multisig}, \ref{sec_snark_agg}, and \ref{sec_hashbased_multisig}, several
constructions in those subsections are forward secure under their stated synchronized or
stateful models, and several others can be lifted into a forward secure variant by
sum composition with a key evolution tree. This subsection collects the constructions whose
forward security profile is the central property, and points back to constructions whose
forward security was understated above.

\paragraph{Sum composition / Key Evolving Signatures (KES).}
Malkin, Micciancio and Miner \cite{mmm-kes} introduced sum composition, a binary tree of
one time signing keys in which the root certifies child keys and each leaf is used exactly once
before being erased. Ouroboros Praos and Cardano deploy MMM style KES on Ed25519 leaves in
production for block producer signatures \cite{cardano-kes}. The construction is the most
operationally mature forward secure signature deployment in any blockchain. KES is agnostic to
the leaf signature, substituting an ML-DSA, Falcon, HAWK, or hash based leaf primitive yields a
forward secure post quantum block producer signature with state size logarithmic in the total
number of epochs. Q10 holds by construction. Q4 to Q5 inherit from the leaf scheme, in
particular, MMM KES on a leaf that cannot aggregate does not by itself provide an aggregate quorum
certificate, it provides forward security for individual validator votes, which must then be
aggregated by a separate Q class construction. A KES based QC claim must therefore be checked
under Q1$'$ rather than by separately claiming Q1 and Q10.

\paragraph{Pixel as pre quantum baseline.}
Drijvers, Gorbunov, Neven and Wee construct Pixel \cite{pixel}, a pairing based forward secure
aggregate signature with key evolution and BLS style non interactive aggregation. Pixel
simultaneously satisfies Q2, partial Q5, and Q10 in the pre quantum setting, and is the correct
operational anchor for ``what a forward secure aggregable signature looks like.'' No published
post quantum construction matches Pixel's simultaneous Q class profile. We treat the
post quantum Pixel analog as the principal open problem of the forward secure aggregation line,
see Section~\ref{sec_open}.

\paragraph{Squirrel and Chipmunk, reframed.}
The synchronized lattice multi signatures of Section~\ref{sec_lattice_multisig} are not merely
``synchronized, bounded time'', the synchronized model \emph{is} the forward security
mechanism. Each signing time step has its own short lived key, the chain's notion of time
drives key evolution, past signing keys are erased, past step compromise is not retroactively
dangerous. Within the bounded $2^\tau$ horizon, Q10 holds. The relevant deployment question
is the choice of $\tau$ at genesis (or a re keying ceremony at $\tau$ expiry) and the size of
per step state retained for slashing window evidence.

\paragraph{DKKW/LeanSig as implicit forward secure constructions.}
The hash based multi signature line of Section~\ref{sec_hashbased_multisig} inherits forward security
from the underlying XMSS style leaf burning structure, once a leaf is consumed, the
corresponding signing capability is irrecoverable. This is one of the strongest arguments for
accepting the state management overhead of hash based validator signatures. State management
has value because it buys Q10 in addition to whatever post quantum security the construction
inherits from hash assumptions. The composed question is whether the aggregate and evidence
layers preserve Q1$'$ when validators crash, roll back, or restore stale signing state.

\paragraph{Operational evaluation.}
Against the QC functionality the KES family satisfies Q1 to Q2 under leaf signature assumptions, Q9,
and Q10, but not Q4 to Q5 natively. Pixel satisfies Q1 to Q5 partially and Q10 fully in the
pre quantum setting. Squirrel and Chipmunk satisfy Q1 to Q3, partial Q4 to Q5, and Q10 within
their synchronized horizon. DKKW/LeanSig achieves the strongest combination currently
published in the post quantum setting, Q1$'$, Q2 to Q4, partial Q5 to Q6, Q7 to Q9, and Q10.

\paragraph{Verdict.}
For chains where Q10 is a hard requirement, the practical post quantum options today are
KES on PQ leaf (simple, no aggregation), Chipmunk (synchronized aggregation, bounded horizon),
or DKKW/LeanSig (closest to BLS interface, larger aggregates). Pixel sets the operational bar
that no PQ construction currently meets.

\subsection{Direct lattice multi signatures}
\label{sec_lattice_multisig}

A direct multi signature avoids SNARKs, the aggregate is a small lattice element built from the
component signatures via algebraic combination, and verification is a single lattice arithmetic
check. Three constructions are particularly relevant.

\paragraph{Boneh to Kim aggregate signatures from lattices.}
Boneh and Kim \cite{boneh-kim-agg} construct aggregate signatures from lattice assumptions with
explicit caps on the number of aggregable signatures, parameter sets supporting up to $N=32$ in
one regime and up to $N=1024$ in another. Within those caps, the construction is
non interactive and public. Q5 is partially supported in the sense that adding one signature to
an existing aggregate is cheap, but only up to the cap and only under the construction's merge
semantics, exceeding it requires a separate aggregate. Q4 holds within a parameter regime but
verification is not as cheap as BLS multi pairing. Q6 is achievable only if the profile defines
admissible evidence or retained openings. The hard cap on $N$ is the operational issue because
rotating committees with hundreds to thousands of validators do not naturally fit within the
parameter regimes that yield small aggregates.

\paragraph{Squirrel.}
Squirrel \cite{squirrel} is a synchronized lattice multi signature, signing takes a time
parameter as input, and aggregation requires that all signatures agree on that time step.
Within the synchronized model, Squirrel achieves non interactive aggregation of up to $\rho$
signatures per step over $2^\tau$ time steps. The synchronized model is a natural fit for
blockchain settings where time is provided by the chain itself. Aggregate sizes are in the tens
of kilobytes for committee scale aggregation. Q5 holds partially within a time step. Q6
evidence is achievable from per signer artifacts retained during the slashing window. Q7
requires care, rotation across $2^\tau$ time steps is supported by construction, but the
parameters are fixed at setup, so a long term deployment must commit to a $\tau$ at genesis or
accept periodic re keying ceremonies.

\paragraph{Chipmunk.}
Chipmunk \cite{chipmunk} improves on Squirrel, smaller aggregate sizes, better signing
performance, and explicit rogue key resistance via homomorphic vector commitments. Like
Squirrel, Chipmunk is synchronized and bounded time, the same Q7 considerations apply. In the
synchronized blockchain setting, Chipmunk is arguably the strongest direct lattice
multi signature currently available.

\paragraph{MuSig-L and TOPCOAT.}
MuSig-L \cite{musig-l} is a lattice multi signature with a single online round but
\emph{interactive} aggregation, signers must communicate during signing. This fails Q2
(public non interactive aggregation). TOPCOAT~\cite{topcoat} is a two party Dilithium
signature, which is a threshold construction rather than a multi signature and addresses a
different setting. Neither is a direct candidate for the QC role as we have specified it.

\paragraph{Verdict for the family.}
Within the synchronized model, Chipmunk is the most compelling currently published direct
lattice multi signature for a blockchain QC role. Its limitations are the bounded time
horizon, the non-trivial setup complexity, the need to evaluate Q1$'$ rather than Q1/Q10
separately, and aggregate sizes of tens of kilobytes versus BLS's 48 to 96 byte aggregate
signatures.

\subsection{Half aggregation}
\label{sec_half_agg}

Half aggregation is a distinct operating point between individual individual signatures and
full SNARK style aggregates. A half aggregate is a concatenation of component signatures plus a
shared challenge or transcript that allows batched verification with a constant factor speedup,
without producing a sublinear size certificate. Aggregate size remains $\Theta(|S|)$ but
cryptographic verification cost is meaningfully below $|S|$ independent verifications, and the
security analysis is substantially simpler than for SNARK style aggregation.

Boudgoust and Takahashi \cite{boudgoust-takahashi-halfagg} construct sequential half aggregation
for lattice signatures, with follow up work extending the approach to Falcon like and
Dilithium like schemes. The resulting half aggregate verifier is faster than independent
verification in the cryptographic component, with the linear cost dominated by reading the
component signature blobs.

Against the QC functionality, Q1 to Q3 hold under the underlying signature assumptions, Q4 is
\emph{not} satisfied in the strict sense because aggregate size and descriptor are linear in
$|S|$, Q5 is more natural than for SNARK style aggregates because fresh addition and disjoint
merge are concatenations under a derived again shared challenge, and overlap handling reduces to
descriptor deduplication, Q6 accountability is automatic because individual component
signatures are present in the aggregate. Q11 fails for committee scale aggregation, Q12 is
favorable.

\paragraph{Verdict.}
Half aggregation occupies the right slot for chains where bandwidth budgets tolerate
linear in $|S|$ certificates but verification budgets do not tolerate $|S|$ full verifications.
It is operationally compatible with accountability, although it fails QBLS like and QHT size
budgets.

\subsection{SNARK style aggregation}
\label{sec_lattice_snark}
\label{sec_snark_agg}

A SNARK style aggregate is a succinct argument that a list of individual signatures is valid,
replacing per signer verification with verification of a small proof. This pattern is generic
over the underlying signature scheme and proof system. We consider three instances.

\paragraph{LaBRADOR aggregated Falcon.}
LaBRADOR style aggregation \cite{labrador} is a succinct lattice argument with proof size
polylogarithmic in the number of aggregated relations. Aardal, Aranha, Boudgoust, Kolby and
Takahashi \cite{aardal-labrador} provide a rigorous treatment of aggregating Falcon signatures
with LaBRADOR, including a knowledge soundness analysis for the non interactive version and the
predicate special soundness framework. The construction is the leading post quantum
SNARK style aggregation proposal.

Against the QC functionality, the construction has clear strengths and clear gaps. Q2 (public
aggregation) and Q4 (sublinear cryptographic verification) hold by construction. The gaps are
operational. First, standard LaBRADOR aggregation is not algebraically incremental in the BLS
sense, adding one late vote changes the public statement and normally requires proving the new
statement, rather than performing a constant time group operation on an existing aggregate.
This breaks Q5 in any consensus protocol where votes arrive asynchronously and partial
certificates are gossiped. Second, signer set and slashing evidence (Q3 and Q6) must be
designed explicitly, if the public statement lists every signer and vote transcript, then two
accepted aggregate proofs over conflicting messages constitute slashing evidence under the
soundness theorem, but if the chain needs small per validator slashing evidence the aggregate
must provide local openings or the system must retain component signatures for the slashing
window. Third, LaBRADOR Falcon inherits the Falcon signing caveats from Section~\ref{sec_falcon}, a
single trusted aggregator cannot fix cross platform signing reproducibility for the underlying
validator signers. Fourth, the recursive Fiat to Shamir parameter regime for non interactive
LaBRADOR demands careful parameter choice across the recursion depth, the knowledge error bound
degrades multiplicatively with depth, so committee size changes may require parameter retuning.

\paragraph{HAPPIER.}
HAPPIER \cite{happier} is a hash based aggregatable signature implemented via the Risc0
zero knowledge virtual machine. The aggregator runs the verification of a hash based signature
inside a SNARK over a generic VM, producing a succinct proof of aggregate validity. The
advantage over LaBRADOR Falcon is genericity, any hash based signature can be aggregated in this
style. The disadvantages are the prover overhead of running a zkVM based proof and the
inheritance of the underlying SNARK's trust model. Because HAPPIER is implemented through a
zkVM, the blockchain profile must also specify the proof system, recursion or composition
mode, trusted setup or transparent setup assumptions, VM version, and verifier bytecode as
consensus objects. Otherwise the signature scheme has been replaced by a moving SNARK
implementation target. Q class properties are similar to LaBRADOR Falcon, Q2 and Q4 hold,
while Q5 remains the same gap.

\paragraph{Folding scheme aggregates.}
Folding schemes, including Nova style systems, HyperNova, KZH fold \cite{kzh-fold} and
successors, suggest a path to incremental SNARK style aggregation, each new vote incrementally
extends the proof without proving again the prior statement. KZH fold in particular is presented
as accountable voting from sublinear accumulation, which directly targets the QC setting. There are several proposals for folding schemes built from lattices such as Lova, LatticeFold+, and recently ProtogaLattice \cite{lova,latticefold,protogalattice}. Lattice based folding schemes generally rely on commitments to some short lattice vector. Naively, we can combine, or fold, two commitments into one by taking a random linear combination of them. However, we cannot generally claim that the underlying folded witness is still short, even if the commitments are individually. Lova solves this problem by decomposing the witnesses using a gadget matrix, then commits to the decompositions and takes a random linear combination to fold them. By careful selection of parameters, this approach ensures that the norm of the folded witness does not grow too fast. A similar approach exists in code based cryptography, \cite{lpn-aggregate} proposes secure aggregation based on LPN where the Hamming weight of the aggregate errors is kept within the decoding threshold by decomposing the errors using a sparse basis. ProtogaLattice~\cite{protogalattice} uses a bootstrapping argument to reduce the norm of folded witnesses instead.
We
are unaware of a complete published construction that closes Q5, including overlap safe merge
under gossip, for a deployed post quantum signature scheme. Even so, the direction is the most
plausible route to a SNARK style aggregate that matches the BLS operational interface. We treat
this as a leading open problem (Section~\ref{sec_open}).

\paragraph{Verdict for the family.}
SNARK style aggregation gives the smallest aggregate certificates among post quantum proposals
and the most flexibility in choice of underlying signature, at the cost of Q5 incompatibility
with current constructions, larger prover cost, inherited SNARK trust model considerations, and
explicit Q6 evidence design. Folding scheme aggregates are the natural next step but are not
yet production ready BLS replacements.

\subsection{Hash based multi signatures (DKKW, LeanSig)}
\label{sec_hashbased_multisig}

The most operationally complete post quantum BLS replacement line currently published is the
hash based multi signature framework of Drake, Khovratovich, Kudinov and
Wagner \cite{drake-hashbased-mulsig}, with refinement in LeanSig \cite{leansig}. This work was
developed explicitly for Ethereum's consensus migration and explicitly targets the operational
interface of BLS aggregation rather than a generic aggregate signature notion.

The construction generalizes XMSS style hash based signatures with tweakable hashes and
incomparable encodings. Per validator vote signatures are in the 2 to 5 KB range. Aggregate
quorum certificates are parameter dependent and are better described as tens of kilobytes than
as 2 to 5 KB objects. Verification of the aggregate certificate is sublinear in the committee
size under the construction's succinct argument layer. The underlying primitive is hash only,
which keeps the base signature assumption minimal and avoids both Falcon's floating point
signing surface and the lattice assumption maturity question of HAWK.

Against the QC functionality, Q1 and Q2 hold under the stated hash based and proof system
assumptions, for key evolving deployments Q1$'$ is the relevant claim, Q3 holds when the signer
descriptor and committee root are committed in the statement, Q4 is the main advantage, Q5 is
partially supported and is closer to the BLS operational interface than ordinary SNARK
aggregation, but it is not algebraically identical to BLS group multiplication, Q6 is best
marked partial until a formal accountability theorem specifies whether full certificates,
retained component signatures, or local openings are the canonical slashing evidence, Q7 is
plausible because epochal rotation is a natural usage pattern, Q8 is simpler than in
pairing based aggregate signatures, and Q9 is favorable because the underlying operations are
deterministic hash computations.

\paragraph{Stateful signing risk.}
Hash based validator signatures move risk from algebra to operations. Any XMSS like or
one time key hierarchy must specify how validators prevent state rollback, duplicate one time
key use, HSM snapshot reuse, crash recovery divergence, equivocation under restored backups,
and slashing window evidence loss. A blockchain deployment should treat these as consensus
requirements rather than wallet UX details. The profile must say which state is slashable,
which state is recoverable, how epoch rotation consumes or retires signing leaves, and which
artifacts must be retained by validators, aggregators, or the chain.

\paragraph{Tradeoffs versus BLS\@.}
The DKKW/LeanSig line does not match BLS on aggregate size, a hash based aggregate is typically
tens of kilobytes versus BLS's 48 to 96 byte aggregate signatures. It does, however, match more
of the rotating validator operational interface than ordinary lattice or generic SNARK style
aggregation.

\paragraph{Verdict.}
The hash based multi signature line is the most credible currently published candidate for a
post quantum QC primitive that targets BLS style consensus operations. It is the natural
default for a chain that prioritizes interface compatibility, Q1$'$ forward secure soundness,
and assumption conservatism over aggregate byte size. Where aggregate size dominates,
LaBRADOR style constructions remain the smaller alternative at the cost of Q5.

\section{Failure matrix}
\label{sec_failure_matrix}
The following table is deliberately harsher than a primitive survey because it records
why a construction fails as a drop in blockchain replacement even when it has a
credible raw security story.

\begin{table}[H]
\centering
\scriptsize
\setlength{\tabcolsep}{3pt}
\renewcommand{\arraystretch}{1.08}
\begin{tabularx}{\textwidth}{lYYY}
\toprule
\textbf{Scheme/profile} & \textbf{transaction side blocker} & \textbf{QC side blocker} & \textbf{bottom line} \\
\midrule
ML-DSA & fails TMTU/THT byte budgets, multi kilobyte signatures & no native aggregation & conservative TGEN primitive, outside high throughput and QC use \\
SLH-DSA & fails transaction rate byte budgets & no native aggregation & root key primitive, outside the hot path \\
Falcon/FN-DSA & strict encoding and signing profile required, strong UF and AWCV profile must be explicit & no native aggregation, inherits Falcon signing issues under SNARK aggregation & compact with operational fragility \\
HAWK & younger assumptions, compressed decoding AWCV not yet blockchain profiled & no native aggregation & promising transaction candidate, not yet a conservative default \\
MAYO & younger multivariate security and missing AWCV profile & no native aggregation & worth watching, not yet a default account primitive \\
SNOVA R2 & submitted parameters attacked / in flux & no native aggregation & should not be treated as deployable \\
UOV / QR-UOV & public keys awkward for account storage or tx inclusion & no native aggregation & functionally valid with poor account model fit \\
CROSS & eliminated after Round 2, signatures too large for high rate transaction lanes & no native aggregation & historical code based diversification candidate \\ FAEST & signatures and verification too heavy for high rate transaction lanes & no native aggregation & active Round 3 low frequency diversification primitive \\ MQOM / SDitH & several kB signatures and missing blockchain profiled AWCV & no native aggregation & active Round 3 candidates, without compact transaction or QC coverage \\
SQIsign & signing latency and side channel hardening incompatible with hot path signing & no native aggregation & compact archival / low frequency primitive \\
Half aggregation & linear certificate size & fails strict Q11 at large $n$ & useful when verification, not bandwidth, is the bottleneck \\
Squirrel / Chipmunk & not a transaction primitive & bounded synchronized horizon, tens of kB aggregates, partial merge semantics & best direct lattice QC line, still not BLS like \\
LaBRADOR Falcon & not a transaction primitive & fails Q5 incrementality, Q6 needs retained signatures or openings, prover cost & succinct without gossip native merge \\
HAPPIER & not a transaction primitive & zkVM/SNARK trust model and prover cost, fails Q5 incrementality & generic without BLS replacement semantics \\
DKKW / LeanSig & validator specific, stateful & tens of kB certificates, state rollback and Q6 evidence policy must be formalized & closest published PQ BLS interface candidate, still not BLS like \\
\bottomrule
\end{tabularx}
\caption{Blocking failures. This table is deliberately harsher than the functional matrix because it records why a construction is not a drop in blockchain replacement even when it satisfies several individual requirements.}
\label{tab_blocking_failures}
\end{table}

\section{Blockchain case studies}
\label{sec_case_studies}

The requirements outlined in this paper are intentionally abstract, but the pressure points become clearer when instantiated in concrete systems. This section does not propose migrations for the chains below. Rather, we use them as stress tests for the requirements. We give a more thorough treatment for Bitcoin and Ethereum as the two most dominant chains and for Sei Giga as an example of a next generation, high throughput EVM/BFT stress profile.

\begin{table}[!htbp]
\centering
\scriptsize
\setlength{\tabcolsep}{1.5pt}
\renewcommand{\arraystretch}{0.98}
\begin{tabularx}{\textwidth}{lYYYY}
\toprule
\textbf{System} & \textbf{current signature shape} & \textbf{PQ bottleneck} & \textbf{requirement stress} & \textbf{research gap exposed} \\
\midrule
Ethereum & EOA transactions plus BLS attestations and sync committees & transaction byte policy and BLS interface replacement & T3 to T11, Q2 to Q6/Q10 to Q13 & accountable PQ aggregation for attestations and light clients \\
Tendermint/CometBFT & proposal, prevote, and precommit vote signatures & linear commits only scale to modest $n$ and block rates & Q3, Q6, Q11, Q12 & evidence preserving compression for signed votes \\
Algorand & Ed25519 accounts, participation keys, VRF selected committees, and an AVM Falcon verifier & native PQ account profiles plus PQ consensus selection/signing machinery & T3 to T11, adjacent VRF and key evolution issues & strict Falcon/FN-DSA profiles and PQ VRFs \\
Sui & user signatures with scheme flags plus BLS authority certificates and Mysticeti signed block consensus & PQ account agility, authority certificate replacement, and signed DAG/block costs & T3 to T11, Q2 to Q8/Q11 to Q13, fast path evidence & PQ replacement for BLS authority signatures and fast path certificates \\
Solana today & MTU limited Ed25519 transactions, first signature is transaction id, validator votes are transaction like hot path objects & PQ signatures do not fit the 1232 byte packet model, txid policy conflicts with randomized PQ signing & T4, T8, T9, T10, T11 & compact PQ authorization or transaction format redesign \\
Bitcoin & ECDSA and BIP340 Schnorr UTXO authorization, no validator QC & block weight, witness policy, and legacy UTXO migration & T3, T8, T9, downgrade resistance & compact PQ witness programs and rescue rules \\
Next generation BFT / Sei Giga & EVM authorization plus high rate BFT certificates & line rate signature bytes and fast mergeable QCs & T4, T9, T10, T11, Q5, Q11, Q12, Q13 & compact tx signatures and BLS like PQ certificates \\
\bottomrule
\end{tabularx}
\caption{Case study summary. Solana is separated from next generation BFT because today's Solana is primarily an MTU limited transaction signature problem, whereas a future certificate based design becomes a Q class problem.}
\label{tab_case_study_summary}
\end{table}

\begin{observation}[Linear certificate pressure w/o aggregation/proofs]
\label{obs_linear_qc_pressure}
Fix a committee vote transcript and a post quantum signature profile whose accepted signatures
have length at least $L_{\sig}$. Any transparent quorum certificate format that contains no
aggregate algebra, no succinct proof, and no local opening mechanism, but that carries
signer level evidence inside the certificate for every signer in $S$, has size
$\Omega(|S|\cdot L_{\sig})$ up to descriptor overhead. If those component signatures are not
inside the certificate, then Q6 accountability is being supplied by an external retention
policy rather than by the certificate bytes themselves. This is a definitional pressure point rather than
a new lower bound. Under those restrictions the component signatures are the evidence, so
avoiding the linear byte cost requires exactly one of the mechanisms excluded in the statement.
\end{observation}

\begin{table}[H]
\centering
\scriptsize
\setlength{\tabcolsep}{2.2pt}
\renewcommand{\arraystretch}{1.08}
\begin{tabularx}{\textwidth}{lYYY}
\toprule
\textbf{Stress case} & \textbf{Compiled requirement set} & \textbf{Schemes that fail first} & \textbf{Research target} \\
\midrule
Bitcoin class UTXO migration
 & T2 to T4, T8 to T11, T17, H1 to H6, no Q class requirement unless a future covenant or federation layer adds committee certificates
 & SLH-DSA and ML-DSA fail compact spend economics first, Falcon/HAWK need strict witness encodings and invalid input metering, SQIsign fails hot wallet signing latency
 & A soft forkable PQ witness program with below one kilobyte signatures, no txid malleability, and explicit legacy output rescue/sunset policy. \\
Ethereum class PoS migration
 & T1 to T17 for accounts and account abstraction, Q1 to Q18 for attestations, sync committees, finality proofs, and light clients, H1 to H6 during transition
 & Single signer ML-DSA/Falcon/HAWK/MAYO fail Q2 to Q6 natively, LaBRADOR Falcon fails Q5, Chipmunk/Squirrel are too large and synchronized, DKKW/LeanSig pays tens of kB QCs and state discipline
 & A PQ BLS interface replacement with public aggregation, accountable openings, forward secure soundness, and decentralized gossip merge. \\
Sei Giga / high throughput BFT stress profile
 & T4, T8 to T11, T15, T17 plus Q2 to Q6 and Q11 to Q15 at below one second cadence, light client and bridge verification inherit L/X budgets
 & ML-DSA fails line rate transaction bytes, SLH-DSA fails hot path entirely, Falcon/HAWK/MAYO remain plausible only with AWCV artifacts, all current PQ QCs fail QBLS like size/merge targets
 & A split profile, compact transaction authorization below roughly one kilobyte and accountable QCs with fresh add, disjoint merge, overlap safety, and low prover latency. \\
\bottomrule
\end{tabularx}
\caption{The three main stress profiles. The table is the practical reading of the compiler, a primitive that is acceptable for one row can be irrelevant or failing for another.}
\label{tab_three_stress_reqs}
\end{table}

\subsection{Bitcoin}
\label{sec_case_bitcoin}

Bitcoin is the cleanest example of a chain whose primary signature problem is transaction
authorization rather than quorum certificates. Bitcoin has no validator committee and no
BLS style aggregate QC. Its current signature layer is ECDSA plus BIP340 Schnorr over
secp256k1, BIP340 specifies fixed 64 byte Schnorr signatures, 32 byte x only public keys,
byte level verification, tagged hashing, key prefixing, batch verification considerations, and
malleability resistance through SUF-CMA security \cite{bip340}. SegWit moved witness data, including
scripts and signatures, outside the legacy transaction hash, introduced a separate witness
transaction identifier, and defines block weight as base size times three plus total size with
a 4,000,000 weight unit block limit \cite{bip141}. A 2026 draft informational BIP proposes a
post quantum migration and legacy signature sunset path, illustrating that Bitcoin migration is
already being discussed as a protocol policy problem rather than a library replacement
\cite{bip361}.

For a pure upper bound calculation, assume the post quantum signature is carried entirely in
witness data. Under BIP 141, block weight is
\[
 3\cdot\mathsf{base\_size}+\mathsf{total\_size},
\]
so one witness byte contributes one weight unit, while one non witness byte contributes four.
Ignoring all non signature data, a 4,000,000 weight block could therefore contain at most
\[
\frac{4{,}000{,}000}{2420}\approx 1650
\]
ML-DSA-44 witness signatures,
\[
\frac{4{,}000{,}000}{666}\approx 6000
\]
Falcon-512 padded witness signatures, or
\[
\frac{4{,}000{,}000}{7856}\approx 509
\]
SLH-DSA-128s witness signatures. If the same bytes were carried in non witness data, each
bound would be divided by four. These are generous upper bounds, public keys, scripts, outputs,
input metadata, annex data, control blocks, and fee market behaviour all reduce the actual
count.

Bitcoin also has a distinctive public key exposure problem. Hash to public key output types
hide the public key until spend, while other outputs expose public keys earlier or permanently.
A migration profile must therefore specify a new PQ witness program together with what
happens to legacy outputs, already exposed public keys, lost keys, and timelocked or covenant
constructions. This is the sharpest version of Section~\ref{sec_migration}, the chain cannot simply
add a new algorithm identifier and wait.

\paragraph{Bitcoin benchmark problem.}
A useful benchmark is a soft forkable PQ witness program with below one kilobyte signatures,
consensus deterministic decoding, transaction id malleability resistance compatible with SegWit, and
a concrete migration path for legacy keys. Aggregate signatures are secondary here, compact
single spend authorization and upgrade incentives are primary.

\subsection{Ethereum}
\label{sec_case_ethereum}

Ethereum is the hardest case because it uses signatures in both transaction and quorum certificate roles at once. The execution
layer has user transaction authorization and smart contract visible signatures. The consensus
layer is built around BLS, beacon blocks contain BLS signatures, attestations carry an
\texttt{aggregation\_bits} descriptor and a \texttt{BLSSignature}, indexed attestations require
sorted unique validator indices and a BLS fast aggregate verification call, and Altair adds a
512 validator sync committee whose block body object contains a bitvector plus a single BLS
aggregate signature \cite{eth-consensus-phase0,eth-consensus-altair,eth-altair-bls}. Attester
slashing is expressed as two indexed attestations whose data conflict under the Casper FFG
slashing relation \cite{eth-consensus-phase0}.

\paragraph{Current Ethereum PQ policy environment.}
Ethereum's current public PQ roadmap makes this case study more concrete than a
generic BLS replacement exercise. The execution layer path is framed around
account abstraction, PQ signature precompiles, gradual opt in migration, and
eventual PQ transactions. The consensus layer path is framed around replacing
BLS validator signatures with hash based signatures such as leanXMSS, and using
leanVM or related SNARK machinery to recover aggregation efficiency
\cite{pq-ethereum,eth-quantum-resistance-page}. Vitalik's quantum emergency
thread also makes the account exposure issue explicit, once an ECDSA public key
has been exposed by a transaction, a sufficiently strong quantum attacker can
target the corresponding private key, whereas never used address hashes have a
different exposure profile~\cite{buterin-quantum-emergency}.

This policy environment strengthens the requirements in the account abstraction and credential placement requirements. Ethereum PQ transactions are not merely a
choice among ML-DSA, Falcon, and SLH-DSA. They require a decision about native
account abstraction, smart contract wallet verification, residual ECDSA wrapper
transactions, public key recovery versus public key storage, and whether
transaction or mempool aggregation is worth its prover cost
\cite{sanso-pq-aa,sanso-thiery-wagner-pq-mempools}. On the consensus side,
recent Ethereum Research discussion of folding and recursive aggregation
correctly treats the hard problem as decentralized aggregation over overlapping
sets of validator signatures rather than a one shot SNARK over a fixed batch
\cite{eth-pq-aggregation-folding}.

The execution layer migration is a T class problem. It needs canonical public key and
signature bytes, a transaction hash policy, hybrid combiner rules, and worst case gas accounting
for invalid inputs. A smart contract ecosystem also inherits a second metering problem, if PQ
verification is exposed as a precompile or opcode, the gas price must be calibrated to malformed
public keys and malformed signatures, not merely to valid KATs. This is the EVM version of
T4/T11.

The consensus layer migration is a Q class problem. A naive replacement of each BLS aggregate
by individual ML-DSA signatures is not close. Ethereum's phase 0 preset targets committees of
size 128 and allows up to 128 attestations in a block \cite{eth-consensus-phase0}. A deliberately
rough upper bound calculation gives
\[
128\cdot 96 = 12{,}288 \text{ B}
\]
for the BLS aggregate signature bytes in 128 attestations, while replacing 128 signers per
attestation by ML-DSA-44 signatures would cost
\[
128\cdot 128\cdot 2420 \approx 39.6 \text{ MB}
\]
before bitlists, public keys, attestation data, and other block contents are counted. This is
a stress calculation rather than a claim about Ethereum's average block payload, and it shows why
single signer PQ migration does not solve Ethereum consensus.

Ethereum also makes Q6 non negotiable. The current BLS profile gives compact aggregate
attestations and separate indexed attestation objects for slashing. A PQ replacement that only
proves ``some quorum signed'' is not enough, it must identify the accountable validator
indices, bind the exact fork/epoch/slot/\\domain transcript, and produce evidence accepted by
full nodes and light clients. This is why the DKKW/\\LeanSig line is interesting despite its
large certificates, it targets the consensus interface rather than only the byte size of a
single signature.

\subsection{Sei Giga}
\label{sec_case_seigiga}

Sei Giga serves as a representative of next-generation high-throughput BFT blockchains, providing a public stress profile that captures the stringent latency and communication requirements of emerging blockchain architectures. The public whitepaper
describes a multi proposer, parallelized execution EVM layer 1 with Autobahn consensus,
consensus over transaction ordering before asynchronous execution, no traditional mempool,
and reported internal testnet targets above 5 gigagas/s, more than 200K simple transfer TPS,
and below 400 ms finality \cite{sei-giga-paper}. We use those claims only to define an
aggressive public envelope for PQC requirements.

At these rates, transaction signatures are no longer background overhead. At a 200K TPS
engineering target, signature payload alone is approximately
\[
200{,}000\cdot 65 \approx 13 \text{ MB/s}
\]
for compact EVM style ECDSA authorization bytes,\footnote{The 65 byte figure here is the
EVM style compact $(r,s,v)$ authorization form relevant to an EVM compatible transaction
profile. The calibration table reports roughly 71 bytes for DER style secp256k1 signatures,
both are pre quantum baselines, and the distinction matters only for consistent byte accounting.}
\[
200{,}000\cdot 666 \approx 133 \text{ MB/s}
\]
for Falcon-512 padded signatures, and
\[
200{,}000\cdot 2420 \approx 484 \text{ MB/s}
\]
for ML-DSA-44 signatures. These are lower
bounds on the signature byte contribution alone rather than full network bandwidth estimates. They explain why a high throughput EVM chain
cares about Falcon, HAWK, MAYO, or future compact signature profiles even if ML-DSA is the
safer finalized default in lower rate settings.

The validator side is equally unforgiving but in a different way. To make the byte pressure
concrete, consider a 100 validator BFT stress profile. A certificate made of individual
ML-DSA-44 signatures is roughly
\[
100\cdot 2420 \approx 242 \text{ KB}.
\]
Even if a chain budgets tens of megabytes for block data, this is expensive for below one second
gossip, light clients, and repeated certificates across ordering, data availability, and
state root attestation. It also illustrates why Q5 and Q6 cannot be afterthoughts. If a
high rate BFT chain publishes only compact certificates, the certificates still need explicit
merge semantics and explicit public evidence for which validators signed which transcripts.

Sei Giga therefore compresses both frontiers simultaneously. T9/T10/T11 force compact, fast,
invalid input metered transaction signatures. Q5/Q6/Q11/Q12/Q13 force mergeable,
accountable, low latency quorum certificates. A chain can tolerate a large PQ transaction
signature or a large PQ certificate in isolation. A 5 gigagas/s, below one second BFT chain cannot
treat both as afterthoughts.

\paragraph{Sei Giga benchmark problem.}
A useful high throughput benchmark is a 100 validator, below one second BFT profile with multiple
certificates per second, EVM compatible transaction admission at 100K to 200K TPS, a tens of MB
block data envelope, transaction signatures below roughly one kilobyte, and PQ certificates
that support fresh add, disjoint merge, accountable evidence, and light client verification.

\subsection{Cross case lessons}
\label{sec_case_lessons}

These cases separate four benchmark families that should be explicit in future post quantum QC work.

\begin{description}[leftmargin=*,style=nextline]
\item[Bitcoin class transaction profile.] The target is compact single spend authorization
under a hard block weight budget, strict consensus parsing, and a conservative upgrade path for
legacy outputs. Q class aggregation is mostly irrelevant.

\item[Ethereum/Sui class consensus profile.] The target is a post quantum replacement for BLS
attestations, sync committees, or authority certificates, public aggregation, compact
descriptors, slashable indexed attestations, light client verification, and forward security
compatibility. Single signer NIST signatures are useful only as leaves.

\item[Solana class MTU profile.] The target is packet sized transaction authorization with
stable transaction identifiers and pre execution signature rejection. Multi kilobyte PQ
signatures fail before any consensus aggregation question arises.

\item[High throughput BFT profile.] The target is simultaneous line rate transaction signing
and low latency BFT certificates. This profile is where Falcon/HAWK/MAYO style compactness,
AWCV bounds, and BLS like Q5 mergeability become first order system requirements rather than
optimization details.
\end{description}

This is the strongest systems argument for more post quantum work for blockchains. Existing
standards answer the question ``which single signer signatures should we trust?'' They do not
answer ``which post quantum object is a transaction, a vote, a certificate, a slashing proof,
and a gossip merge state at the same time?''

\FloatBarrier
\section{Structural observations}
\label{sec_structural}

The matrices in the comparison and blocking failure tables expose
several structural patterns that are not visible when post quantum signatures are evaluated one
at a time.

\paragraph{The transaction signature frontier is genuinely unsettled.}
Among lattice schemes, ML-DSA is the safest finalized choice under our requirements but the
largest, Falcon/FN-DSA is the smallest NIST selected lattice option but is profile fragile
around encoding, strong unforgeability claims, AWCV measurement, and signing reproducibility,
HAWK improves the operational signing story but uses younger assumptions and still needs
compressed decoding metering data. Among non lattice schemes, MAYO offers attractive signature
sizes at younger multivariate assumptions, SNOVA had attractive before attack byte geometry but is
now in flux, SQIsign offers very compact keys and signatures at signing latencies that prevent
high throughput use, SLH-DSA, CROSS, and FAEST are too large or slow for transaction rate use
but useful for low frequency high assurance roles. No single scheme dominates once T8, hybrid
policy, invalid input metering, byte profiles, named size classes, and multi user concrete
security are included.

\paragraph{The quorum certificate frontier has a structural gap.}
No currently published post quantum construction simultaneously matches BLS on small aggregate
size, sublinear cryptographic verification, gossip friendly incremental aggregation,
overlap safe merge, accountable evidence, and forward security. SNARK style constructions
(LaBRADOR Falcon, HAPPIER) give Q4 but fail Q5 under ordinary asynchronous vote arrival.
Direct lattice multi signatures (Squirrel, Chipmunk) give partial Q5 within synchronized and
bounded time models but only partial Q4 and much larger certificates. Hash based
multi signatures (DKKW/LeanSig) are the closest to the BLS operational interface, but their
certificates are tens of kilobytes rather than 48 to 96 bytes and Q6 accountability should be
formalized as an explicit theorem. The remaining gap is structural rather than a matter of
minor optimization.

\paragraph{NIST finalization does not settle the blockchain problem.}
FIPS 204 and FIPS 205 are important because they standardize single signer digital signatures,
yet the hard consensus questions are one layer higher, how invalid inputs are metered, whether
transaction IDs include signatures, whether public keys are stored or transmitted, how hybrid
suite policy is bound, how aggregated signer sets are committed, whether overlaps can be
merged idempotently, and what counts as slashable evidence. These are not implementation
details that can be deferred to wallets or clients. They are part of the safety proof of a
public consensus protocol.

\paragraph{Multivariate schemes separate functional validity from account model fit.}
UOV and QR-UOV should not be marked as failing public key validation merely because their
public keys are large. Functionally, T6 can hold, public key decoding and validation can be
deterministic and metered. The problem is deployment fit. A 12 to 272 KB public key is a poor
match for account systems that store or repeatedly transmit user keys. MAYO is therefore the
multivariate candidate worth considering for transaction signing, because it partially closes
the public key size gap while retaining compact signatures. SNOVA like designs would be
interesting if reparameterized securely, but the submitted Round 2 profile should not be
counted as deployable.

\paragraph{Constant time verification and adversarially metered verification are different
properties.}
The cryptographic engineering literature treats ``constant time verification'' as a side channel
hardening property, the verifier's running time should not depend on secret inputs. Several
post quantum verifiers are constant time in this sense. The blockchain requirement
(the metered verification definition) is different, the verifier's running time on adversarial input should not
depend on attacker choices except through an explicitly priced gas schedule. A scheme can be
constant time in the side channel sense and still admit adversarially expensive invalid
encodings. This is the situation under permissive Falcon and HAWK profiles. Existing
performance studies do not consistently distinguish these two properties, and the AWCV/metering
profile of every onramp candidate is, to our knowledge, understudied.

\paragraph{Signing side reproducibility is the load bearing property for distributed signers.}
T7 (seeded signing reproducibility) appears to be a soft requirement when read in isolation,
but it controls whether a chain can run threshold custody, MPC signing, or any deterministic
recovery validator software without architecture by architecture KAT enforcement as a fork
rule. Falcon/FN-DSA's floating point signing surface and HAWK's optional FMA flag are not
academic concerns. They translate into chain level operational costs that are paid forever once
the chain commits. ML-DSA, SLH-DSA, the multivariate family, CROSS, and FAEST do not pay this
particular cost.

\paragraph{Academic crypto is still mostly solving the wrong blockchain interface.}
This is the harsh but accurate reading of the comparison. The community has produced valuable
post quantum single signer signatures. It has not produced a consensus ready post quantum
replacement for the signature layer of a modern public blockchain. The missing object is not
``a signature with smaller bytes.'' It is an object with canonical consensus bytes,
adversarial input cost bounds, hybrid downgrade resistance, transaction name semantics,
public aggregation, accountable evidence, forward secure key evolution, and gossip friendly
merge behaviour. The gap reaches beyond a minor implementation detail. It is an abstraction failure.

\section{Migration profiles and downgrade resistance}
\label{sec_migration}

Algorithm agility alone does not make a migration plan. A chain profile must say, for every
role and epoch, whether classical only, hybrid, or PQ only authorization is
accepted. The required suite must be inside the signed transcript, since otherwise a
hybrid authorization can be replayed or reinterpreted by a classical only path.
This is especially acute for exposed legacy keys, bridge/admin keys, validator
keys after stake withdrawal, and account abstraction wrappers that still rely on
classical EOAs.

A safe hybrid transaction profile uses AND semantics, the classical component
and the PQ component both verify over a transcript that binds chain id, fork
version, role, account or validator identifier, both public key hashes, and the
required suite policy. The stable transaction identifier should be computed
from the signed intent and suite policy, not from randomized signature bytes. If
signature bytes are included in the transaction name, the combined profile needs
a strong malleability resistance argument for already signed intents. Verification
order, early rejection, and gas charging are consensus rules. Malformed hybrid
authorizations must be charged for the worst admitted path before an attacker can
externalize parsing or verification cost. Hybrid signature design goals such as
nonseparability, proof composability, and strong unforgeability are therefore
blockchain requirements, not merely API preferences~\cite{bindel-hale-hybrid-sigs,ietf-hybrid-sig-spectrum,cfrg-suf-hybrid-sigs}.

The minimum migration artifact is a sunset schedule that states when classical only
transaction signatures, validator votes, bridge attestations, governance
messages, and light client updates stop being accepted. A profile that never
sunsets classical authorization provides compatibility rather than post quantum migration.

\section{Open problems}
\label{sec_open}

The requirements above reduce the blockchain PQ signature problem to a small
set of concrete construction targets.

\begin{enumerate}[label=\textbf{O\arabic*.},leftmargin=*]
\item \textbf{Blockchain domain PQ signatures.} Design schemes whose native
API is a transaction, vote, certificate, evidence object, or light client update,
with canonical bytes and adversarial invalid input cost in the model.
\item \textbf{BLS like PQ quorum certificates.} Construct a public,
non interactive aggregate with small certificates, sublinear verification,
fresh add, disjoint merge, overlap safe merge, signer set binding, and local or
retained slashing evidence. This is the missing Q2+Q4+Q5+Q6 object.
\item \textbf{Compact transaction signatures below the THT/TMTU frontier.}
ML-DSA is conservative but too large for packet limited or 100K to 200K TPS lanes,
Falcon, HAWK, MAYO, and future designs need strict byte profiles, reproducible
signing rules, and AWCV measurements before becoming native transaction
schemes.
\item \textbf{Adversarial worst case verification artifacts.} Every candidate
profile should publish negative vectors for malformed keys, noncanonical
signatures, length maximal invalid encodings, wrong role transcripts, duplicate
aggregate signers, stale committee roots, overlap unsafe merges, and invalid
slashing openings.
\item \textbf{Forward secure accountable aggregation.} The post quantum analog
of Pixel~\cite{pixel} remains open, with Q1$'$, Q2, Q4, Q5, Q6, and Q10
simultaneously, under state rollback, crash recovery, and evidence retention
constraints.
\item \textbf{End to end chain shaped benchmarks.} We need Bitcoin witness,
Ethereum attestation/sync committee, and high throughput BFT/Sei Giga profiles
that report bytes, gas, invalid input rejection, evidence retention, and gossip
merge behavior instead of only average valid verification time.
\end{enumerate}

\section{Conclusion}
\label{sec_conclusion}

Post quantum blockchain signatures should be specified as consensus profiles, not as raw
signature names. The profile must say which bytes are accepted, which transcript is signed,
how invalid inputs are charged, how transaction identifiers are computed, how hybrid downgrade
resistance is enforced, and what evidence an aggregate certificate provides. The case studies
make the separation concrete, Bitcoin mainly stresses compact transaction authorization and
legacy output migration, Solana today stresses MTU constrained transaction authorization and
transaction ID policy, Ethereum stresses a BLS like accountable aggregation interface, Sui
stresses the combination of user signature agility and BLS authority certificates,
Tendermint style chains stress evidence preserving compression of signed votes, Algorand
stresses the separation between VM level PQ verification, native account signatures, and
consensus selection machinery, and next generation high throughput BFT chains such as the Sei
Giga stress profile force line rate transaction bytes together with below one second QC mergeability.
Once these requirements are made explicit, the design space looks different from the
size and speed comparison that anchors most TLS derived discussion.

No current scheme simultaneously dominates the transaction signature requirements once
transaction id policy, byte profiles, invalid input metering, reproducible signing, hybrid
combiner semantics, and multi user concrete security are included. No current
quorum certificate proposal matches the BLS operational interface across the full Q1 to Q13 set.
ML-DSA is the safest finalized transaction primitive subject to its size cost, Falcon/FN-DSA is
the smallest selected lattice option at the cost of a strict profile and operational overhead in
signing, HAWK and MAYO relax different parts of the profile fragility question at the cost of
younger security assumptions, SNOVA illustrates how quickly attractive multivariate byte
profiles can change under new cryptanalysis, SLH-DSA, CROSS, FAEST, and SQIsign occupy
different niche roles. On the QC side, LaBRADOR aggregated Falcon gives small aggregate
certificates but lacks Q5 incrementality, the synchronized lattice multi signature line gives
partial Q5 within bounded time horizons, and the hash based multi signature line of Drake et
al. approaches the BLS operational interface most closely at a substantial size penalty. The
simultaneous achievement of small aggregate size, sublinear verification, accountable evidence,
forward secure composed soundness, and gossip friendly incremental mergeability in the
post quantum setting remains the principal structural open problem.

The uncomfortable conclusion is that current post quantum signature research largely fails the
blockchain interface. It gives us valuable single signer primitives, but it usually stops
before the properties that public consensus needs, adversarial byte metering, transaction name
policy, hybrid downgrade resistance, gossip safe aggregation, signer set evidence, and
forward secure committee rotation. ML-DSA, SLH-DSA, Falcon/FN-DSA, HAWK, MAYO, SQIsign,
LaBRADOR, Chipmunk, and DKKW/LeanSig are all useful components. None is a complete
post quantum replacement for the signature layer of a modern public blockchain. The missing
object goes beyond another table of signature sizes. It is a consensus ready post quantum
signature profile, with negative test vectors, concrete invalid input costs, transaction id
semantics, and accountable quorum certificate behaviour.

We do not advocate a single scheme. We advocate writing the requirements down before choosing
one, then rejecting constructions that do not satisfy the interface a chain actually needs.

\bibliographystyle{alpha}
\bibliography{refs}

\end{document}